\documentclass[pdflatex,sn-chicago]{sn-jnl}

\usepackage{graphicx}%
\usepackage{multirow}%
\usepackage{amsmath,amssymb,amsfonts}%
\usepackage{amsthm}%
\usepackage{mathrsfs}%
\usepackage[title]{appendix}%
\usepackage{xcolor}%
\usepackage{textcomp}%
\usepackage{manyfoot}%
\usepackage{booktabs}%
\usepackage{algorithm}%
\usepackage{algorithmicx}%
\usepackage{algpseudocode}%
\usepackage{listings}%

\theoremstyle{thmstyleone}%
\theoremstyle{thmstyletwo}%

\theoremstyle{thmstylethree}%

\newcommand{\estadoe}{b} 
\newcommand{\estadog}{a} 
\newcommand{\estadoi}{i} 

\newcommand{\kete}{\ensuremath{\lvert \estadoe \rangle}}

\newcommand{\proje}{\ensuremath{\lvert \estadoe \rangle \langle \estadoe \lvert}}
\newcommand{\ketp}{\ensuremath{\lvert \mathbf{p} \rangle}}

\newcommand{\ketephk}{\ensuremath{\lvert \estadoe, \, \mathbf{p} \, + \, \hbar \mathbf{k} \rangle}}

\newcommand{\ketpithree}{\ensuremath{\lvert \mathbf{p} \, + \, \hbar \mathbf{k_1} \rangle}}

\newcommand{\ketpethree}{\ensuremath{\lvert \mathbf{p} \, + \, \hbar \left( \mathbf{k_1} - \mathbf{k_2} \right) \rangle}}

\newcommand{\we}{\ensuremath{\omega_{\estadoe}}}

\newcommand{\ketg}{\ensuremath{\lvert \estadog \rangle}}

\newcommand{\projg}{\ensuremath{\lvert \estadog \rangle \langle \estadog \lvert}}
\newcommand{\wg}{\ensuremath{\omega_{\estadog}}}

\newcommand{\keti}{\ensuremath{\lvert \estadoi \rangle}}

\newcommand{\proji}{\ensuremath{\lvert \estadoi \rangle \langle \estadoi \lvert}}
\newcommand{\wi}{\ensuremath{\omega_{\estadoi}}}

\newcommand{\wip}{\ensuremath{\left(\omega_{\estadoi} + \frac{\lvert\mathbf{p} + \hbar \mathbf{k}_1\lvert^2}{2m\hbar}\right)}}
\newcommand{\wep}{\ensuremath{\left(\omega_{\estadoe} + \frac{\lvert\mathbf{p} + \hbar \left(\mathbf{k}_1-\mathbf{k}_2\right)\lvert^2}{2m\hbar}\right)}}
\newcommand{\wgp}{\ensuremath{\left(\omega_{\estadog} + \frac{\lvert\mathbf{p}\lvert^2}{2m\hbar}\right)}}

\newcommand{\Ce}{\ensuremath{C_\estadoe(t)}}
\newcommand{\Cg}{\ensuremath{C_\estadog(t)}}

\newcommand{\Ceptt}{\ensuremath{C_{\estadoe, \mathbf{p} \, + \, \hbar\mathbf{k}_{\text{eff}}}}}

\newcommand{\Cept}{\ensuremath{C_{\estadoe, \mathbf{p} \, + \, \hbar \left(\mathbf{k}_1 -\mathbf{k}_2\right) }}}

\newcommand{\Cipt}{\ensuremath{C_{\estadoi, \mathbf{p} \, + \, \hbar \mathbf{k}_1}}}
\newcommand{\Cgpt}{\ensuremath{C_{\estadog, \mathbf{p}}}}

\newcommand{\Ceptdot}{\ensuremath{\dot{C}_{\estadoe, \mathbf{p} \, + \, \hbar \left(\mathbf{k}_1 -\mathbf{k}_2\right) }}}
\newcommand{\Ciptdot}{\ensuremath{\dot{C}_{\estadoi, \mathbf{p} \, + \, \hbar \mathbf{k}_1}}}
\newcommand{\Cgptdot}{\ensuremath{\dot{C}_{\estadog, \mathbf{p}}}}

\newcommand{\ketgpt}{\ensuremath{\lvert \estadog, \, \mathbf{p} \rangle}}
\newcommand{\ketept}{\ensuremath{\lvert \estadoe, \, \mathbf{p} \, + \, \hbar \left(\mathbf{k}_1 -\mathbf{k}_2\right) \rangle}}
\newcommand{\ketipt}{\ensuremath{\lvert \estadoi, \, \mathbf{p} \, + \, \hbar \mathbf{k}_1 \rangle}}

\begin{document}

\title[Ultracold Quantum Gravimeters]{Ultracold Quantum Gravimeters: An Introduction for Geophysicists}


\author*[1,2]{\fnm{Ivaldevingles} \sur{Rodrigues de Souza Junior}}\email{ivaldevingles@gmail.com}

\author*[1]{\fnm{Andrea} \sur{Trombettoni}}\email{atrombettoni@units}

\author*[2]{\fnm{Carla} \sur{Braitenberg}}\email{berg@units.it}

\affil[1]{\orgdiv{Department of Physics}, \orgname{University of Trieste}, \orgaddress{\street{Strada Costiera 11}, \city{Trieste}, \postcode{34151}, 
\country{Italy}}}

\affil[2]{\orgdiv{Department of Mathematics, Informatics and Geosciences}, \orgname{University of Trieste}, \orgaddress{\street{Via Weiss 1}, \city{Trieste}, \postcode{34100}, 
\country{Italy}}}


\abstract{This paper aims at providing an accessible introduction to ultracold quantum gravimeters tailored for geophysicists. We do not focus here on geophysical applications, as these are already well known to geophysicists, but rather  provide a pedagogical exposition of the quantum-mechanical concepts needed to understand the operation of quantum gravimeters. We present a review of gravimeters based on two- and three-level atomic systems, focusing on the fundamental mechanisms of atomic interferometry. The functioning of Mach–Zehnder interferometers is discussed through the action of $\pi/2$ and $\pi$ 
pulses, showing how the resulting phase shift encodes gravitational acceleration. The effect of noise is briefly discussed.}


\maketitle

\section{Introduction}\label{sec1}
Gravimeters can be classified into two main categories: absolute and relative \citep{telford1990applied}. Absolute gravimeters determine the absolute value of the gravitational acceleration ($g$) at a specific point on the Earth's surface, whereas relative gravimeters measure only differences in $g$ between two distinct locations.

In this context, spring-based gravimeters are relative instruments, as they require calibration at a site where the value of $g$ is already known, see e.g. \citep{Crossley2013RGReview}. In contrast, the classical absolute gravimeters are based on the principle of free-fall of a test mass in vacuum, whose trajectory is measured interferometrically through a laser with high precision -- e.g. the FG5 gravimeter \citep{Niebauer1995Metrologia}. The absolute instruments are widely employed in the establishment of gravimetric reference stations used for calibrating relative gravimeters \citep{Agostino2007}.

This paper intends to explain the absolute quantum gravimeter starting from the basic derivation of the evolution of the quantum state of the atoms subjected to the gravitational acceleration, so that the interested geophysicist can follow the theoretical framework of the measurement principle. Inherent to the quantum mechanical treatment is the dual property of the atom, having both a mass and a wave nature of the probability amplitude of its position. The mathematical treatment includes the interaction of the atoms with the laser pulses, essential to the gravity measurement. The aim of this effort is to fill the gap between a physicist's description of the quantum gravimeter principle, as discussed e.g. in  \citep{Peters_2001,chu2001, AtomicInterferometry1991,YOUNG1997, Freier_2016}, and the papers describing the applications of the gravimeter as e.g. \citep{Menoret_2018,Wu2019MobileAtomInterferometer}. The former assumes the reader has the background basic quantum physics knowledge obvious for a physicist, the latter have the focus on the observational results. The benefit of the quantum gravimeter is the measurement of the absolute gravity field value, and in principle the absence of drift in the measured values, as explained and 
discussed in the selection of publications explained next.

\citet{Menoret_2018} demonstrated 
the stability and sensitivity of a transportable absolute quantum gravimeter at the level of $10\,\mathrm{nm\,s^{-2}}$ over a 1 month long measurement. \citet{Wu2019MobileAtomInterferometer} illustrated  observations of the tidal and ocean loading gravity signal, with a sensitivity of $20\,\mathrm{nm\,s^{-2}}$. They showed that the effect of the tidal signal is seen in the Allan deviation leading to peaks at long times (starting from several hours and beyond 24 hours) and in the the noise spectrum. The performance of the commercial gravimeter AQG\#B01 of the company Muquans showed absence of instrumental drift over a period of 2 weeks, defined in the limits of the uncertainty range, and repeatability better than $50\,\mathrm{nm\,s^{-2}}$. The performance was tested comparing it to a classical optical interferometric absolute gravimeter (Micro-g-LaCoste, FG5\#228), and a superconducting relative gravimeter (GWR, igrav\#002) in the Larzac observatory in southern France and in a laboratory in Montpellier \citep{cooke2021}. They studied geophysical signals comprised earth tides and hydrological signals, and atmospheric gravity and loading effects. The absence of drift is particularly important for acceleration measurements from space, as the useful spectral measurement bandwidth has no lower frequency limit, as is the case for the electrostatic accelerometer %
\citep{Zingerle2024QuantumAccelerometers,Romeshkani2025QuantumGravimetry, Rossi2023QuantumSat}. The consequence is the increase of the observable spatial wavelength along the satellite orbit, beneficial for the determination of the low degrees in the spherical harmonic expansion of the gravity potential field \citep{Migliaccio2019MOCASS,Migliaccio2023MOCASTPlus,Rossi2023QuantumSat}. The lowered noise spectrum leads further to improved spatial and temporal resolution of the retrieved field with improved resolution of hydrologic, oceanic and solid earth signals \citep{Kusche2025MAGIC, Braitenberg2024MOCAST, Pivetta2022MOCAST}. We refer to 
\citep{Fang2024ClassicalAtomicGravimetry} for a review of classical and atomic static gravimetry as well as airborne/marine, space-based gravimetry.

Essential components of ultracold quantum gravimeter are the ultracold atom source and its excited states, the interaction with three laser pulses, and the atomic interferometric phase difference depending on  gravity. The paper starts with introducing the probability amplitude of an atom, the corresponding Hamiltonian, and then the mathematical derivation of the two- and three-level states of the atom, the phase changes of the probability function introduced by the laser pulses, the phase change acquired along the trajectory in space of the atom, and finally the probability of the atoms to be in the excited state and its relation to the gravity acceleration. The efforts to understand the quantum equations are an intellectual investment which extends beyond the quantum gravimeter.

We observe that the majority of devices for geoscientific applications used in the twentieth century are "classical", i.e., grounded in and exploiting concepts of classical physics. For instance, 
the seismometers used in modern seismology were developed in the early twentieth century, and seismology uses the formulation of the equations governing the propagation of seismic waves in elastic media \citep{LoveSomeProblems,Gutenberg1914}, and, similarly, the magnetotelluric method is based on the principles of electromagnetism \citep{jackson_classical_1999}. From this perspective, up to now a significant part of the academic background needed to work in geoscience remains rooted in classical physics, with few direct applications of quantum mechanical concepts. However, as previously discussed, reliable measurement instruments based on quantum principles that are sufficiently robust and portable for geophysical applications are now more and more becoming available for field and space-based measurements \citep{Freier_2016,Menoret_2018,cooke2021,kanxing_2021,bidel2023airborne,Chen_2023,antonimicollier2024absolute}. Commercial instruments, such as \citep{Glaessel2025ExailAQG}, are gaining applications not only for gravimeter,s but also for inertial navigation \citep{ExailHomepage}. Other devices measure the magnetic field \citep{Rovny2022NanoscaleCovarianceMagnetometry} or the gravity vector in an optical lattice \citep{LeDesma2025}.

The availability and development of quantum gravimeters, and quantum sensors more generally, are part of the broader recent 
progress of quantum science and technologies, which has impacted a wide range of research fields and technological applications — from quantum computing and quantum simulation to quantum sensing. We are now witnessing a significant expansion of their practical applications. The quantum gravimeter, based on atomic interferometry, represents a remarkable example of this new generation of instruments \citep{AtomicInterferometry1991,Menoret_2018}. 

These developments serve as a call to attention for the geoscientific community, encouraging explorations at the intersection between quantum physics and the geosciences. Since, regarding physics, the university background in geoscience and geophysics in bachelor and master formation often (but not always exclusively) has classical physics and  electromagnetism as a core element, we think it is useful to provide an introduction of the essentials of quantum gravimeters which can be helpful for geophysicist readers more accustomed to the classical paradigm and for which many quantum concepts (and, perhaps even more, the notation used  in quantum mechanics) may be rather counterintuitive. For the same reason, we do not discuss the geophysical applications of quantum gravimeters, since they are of course very well known to the geoscience and geophysics communities.

\section{A reminder of quantum mechanical concepts}
The logic we aim in the following is to briefly review some of the fundamental concepts of quantum mechanics essential for their understanding, before introducing the operation mechanisms of a quantum gravimeter. To clarify the notation, while with {\it quantum gravimeter} we refer to a quantum sensor to measure gravity exploiting quantum resources, with {\it ultracold} we indicate quantum gravimeters based on ultracold atoms. We refer to textbooks on ultracold atoms for useful reading \citep{pethick2008,pitaevskii}, here we will mention properties of ultracold atoms when useful for the presentation of properties of ultracold quantum gravimeters, but we decided to rather emphasize the quantum mechanical part of the discussion of their working mechanisms.

\subsection{The superposition principle}

\begin{figure}[H]
	\centering
	\includegraphics[width=0.3\paperwidth]{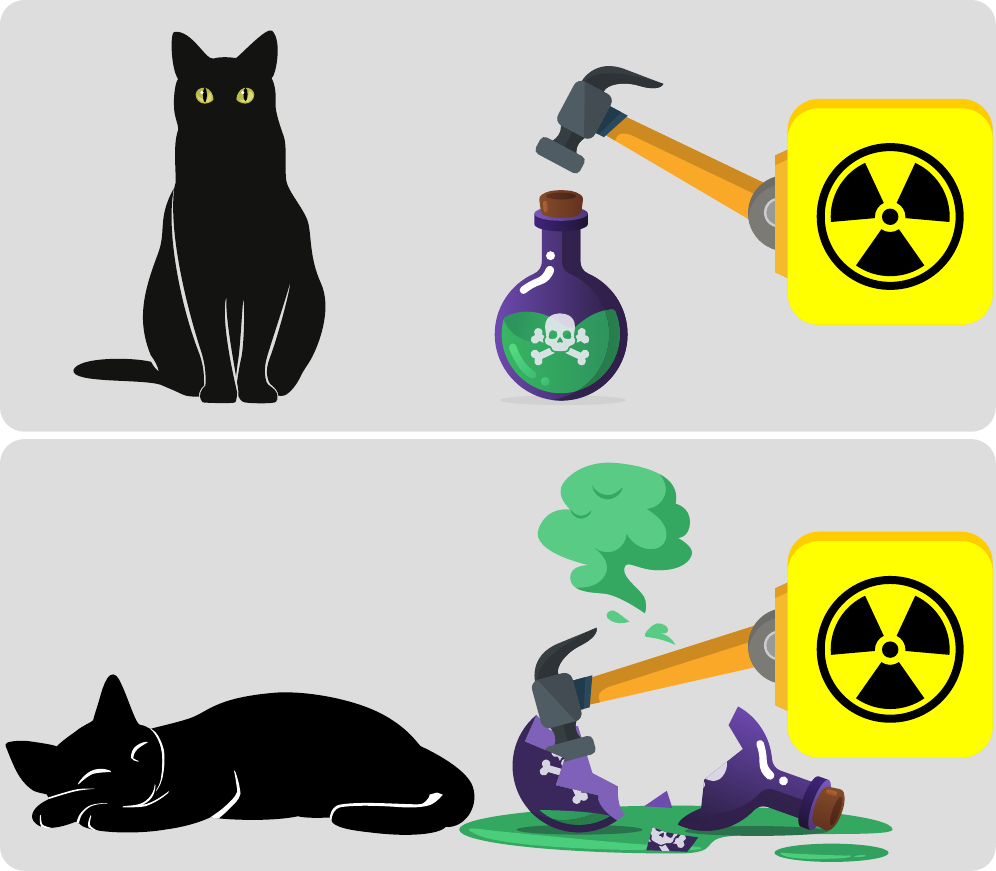}
	\caption{Schematic representation of the 
    Schr\"{o}dinger's cat. In the upper image, the case is illustrated where the radioactive atom has not decayed. In the lower image, the scenario is shown where the atom has decayed, triggering the mechanism that releases the substance, possibly resulting in the death or sleep of the cat.}
	\label{figa:gatovivomorto}
\end{figure}

To illustrate a key mechanism of quantum mechanics, Schr\"{o}dinger proposed a thought experiment that became widely known as the Schr\"{o}dinger's cat 
\citep{Schrodinger1935_English,peres1995quantum}. An original motivation of Schr\"{o}dinger was to discuss the situation in which the laws of quantum mechanics are applied to macroscopic systems, but we do not follow here this discussion path, but rather we use the Schr{\"o}dinger's cat as an example of quantum superposition. The experiment consists of placing a cat inside a sealed box together with: (i) a radioactive atom, which may or may not undergo decay; (ii) a radiation detector, responsible for identifying the decay event; and (iii) a flask containing poison. If the atom decays, the detector activates a mechanism that breaks the flask, releasing the poison, which consequently leads to the death of the cat (Figure~\ref{figa:gatovivomorto}).

According to quantum mechanics, the radioactive atom exists in a superposition of states -- "decayed" and "not decayed" -- until the moment of observation. Since the fate of the cat is directly correlated with the quantum state of the atom, the cat itself is also in a superposition of the states "alive" and "dead". 

Historically the two states are "alive" and "dead", but for kindness to cats presentation reasons, we prefer to think the flask has some substance which can make the cat falling asleep. So, we will talk about "awake" and "sleeping" states. Thus, only when the box is opened and an observation is performed does the system collapse, yielding one of the possible classical outcomes: the cat is either awake or sleeping.

\subsection{The bra-ket notation} 

In the standard formulation of quantum mechanics 
\citep{dirac1930principles,Cohen-Tannoudji:101367,Shankar}, a notation known as the \textit{bra–ket} notation is employed to represent vectors. In this formalism, a generic column vector $(a_{1} \ a_{2} \ a_{3} \ \ldots \ a_{n})^{T}$, with the $a_i$ being complex numbers ($a_i \in \mathbb{C}$), is represented using the \textit{ket} symbol $\lvert \bullet \rangle$ as follows:
\begin{equation}
    \lvert a \rangle =
    \begin{pmatrix}
        a_1 \\
        a_2 \\
        a_3 \\
        \vdots \\
        a_n 
    \end{pmatrix}.
    \label{eq:ket}
\end{equation}

Similarly, the vector corresponding to the conjugate transpose of $\lvert a \rangle$ is called a \textit{bra} and is represented by the symbol $\langle \bullet \rvert$. Thus,
\begin{equation}
    \langle a \lvert =
    \begin{pmatrix}
        a_1^{*} \ \ a_2^{*} \ \ a_3^{*} \ \  \cdots \ \ a_n^{*}
    \end{pmatrix}.
    \label{eq:ket2}
\end{equation}
The bra \eqref{eq:ket} can be written as $\lvert a \rangle = \sum_{i=1}^{n} a_i \lvert i \rangle$, with $\{ \lvert i \rangle\}$ with $i=1,\cdots,n$. This means that $\lvert 1 \rangle=(1 \ 0 \ 0 \ \ldots \ 0)^{T}$, $\lvert 2 \rangle=(0 \ 1 \ 0 \ \ldots \ 0)^{T}$, and so on. The vectors $(1 \ 0 \ \cdots \ 0)^T$, ... $(0 \ \cdots \ 1)^T$ are called basis states -- or canonical basis vectors -- and they form an orthonormal basis of the state space of the system. So therefore writing $\lvert a \rangle = \sum_{i=1}^{n} a_i \lvert i \rangle$ amounts to write the bra $\lvert a \rangle$ in the basis $\{ \lvert i \rangle \}$. In the same way, we have $\langle a \lvert = \sum_{i=1}^{n} a_i^{*} \langle i \lvert$.

The combination of a bra and a ket defines an inner product, whereas the combination of a ket and a bra defines an outer product. Considering the \textit{ket} $\lvert b \rangle$ (with the same dimension as $\lvert a \rangle$), the inner product between these two vectors is given by \citep{Cohen-Tannoudji:101367,Shankar}
\begin{equation}
    \langle a \lvert b \rangle =
    \begin{pmatrix}
        a_1^{*} \ \ a_2^{*} \ \ a_3^{*} \ \  \cdots \ \ a_n^{*}
    \end{pmatrix}\begin{pmatrix}
        b_1 \\
        b_2 \\
        b_3 \\
        \vdots \\
        b_n
    \end{pmatrix} = a_1^{*}\,b_1 + a_2^{*}\,b_2 + a_3^{*}\,b_3 + \cdot + a_n^{*}\,b_n
\end{equation}
and the corresponding outer product takes the form
\begin{equation*}
    \lvert a \rangle\langle b \lvert = \begin{pmatrix}
        a_{1} b_{1}^{*} & a_{1} b_{2}^{*} & \cdots & a_{1} b_{n}^{*} \\
        a_{2} b_{1}^{*} & a_{2} b_{2}^{*} & \cdots & a_{2} b_{n}^{*} \\
        \vdots & \vdots & \ddots & \vdots \\
        a_{n} b_{1}^{*} & a_{n} b_{2}^{*} & \cdots & a_{n} b_{n}^{*}
    \end{pmatrix}.
\end{equation*}

In the bra-ket notation, one can also compute expectation values of observables: if one has an observable ${\cal O}$, corresponding to an Hermitian operator $\hat{{\cal O}}$, one can write it -- in the basis in which the bra \eqref{eq:ket} and ket \eqref{eq:ket2} are written -- as a matrix $O_{i,j}=\langle i \lvert \hat{{\cal O}} \lvert j \rangle$. The expectation value of the observable ${\cal O}$ on a state $\lvert a \rangle$ and is given by $\langle a \lvert \hat{{\cal O}}\lvert a \rangle=\sum_{i,j=1}^{n} a_i^{*} O_{i,j} a_j$.

\subsection{The description of quantum states}

Let us assume that in Schr{\"o}dinger's cat experiment the probability that the cat is awake when the box is opened is 
found to be $\frac{9}{10}$ ($90\%$), and thus the probability that the cat is sleeping is found to be $\frac{1}{10}$ ($10\%$). But how can we represent this ``awake–sleeping" state mathematically? Considering that the quantum state is represented by $\lvert \psi \rangle$, the simplest way to represent the state in which the cat finds itself is by using a column vector that stores the probability of ``observing" each state, that is:
\begin{equation*}
    \begin{matrix}
        \frac{1}{10} \\
        \\
        \frac{9}{10}
    \end{matrix}
    \begin{array}{cl}
        \rightarrow & \text{Probability that the cat is dead/sleeping} \\
                    &                                         \\
        \rightarrow & \text{Probability that the cat is alive/awake}
    \end{array}
\end{equation*}
We can rewrite $\lvert \psi \rangle$ as follows:
\begin{equation}
    \lvert \psi \rangle = \alpha 
    \begin{pmatrix}
        0 \\
        \\
        1
    \end{pmatrix}
    +
    \beta 
    \begin{pmatrix}
        1 \\
        \\
        0
    \end{pmatrix},
    \label{estado2}
\end{equation}
where $\alpha$ and $\beta$ are complex numbers. We can choose one of the two, say $\alpha$, to be real.

The vectors $(1 \ 0)^T$ and $(0 \ 1)^T$ are the basis states and they represent the two possible situations for the cat:
\begin{align}
    \left|\, \raisebox{-1.2em}{\includegraphics[height=3.0em]{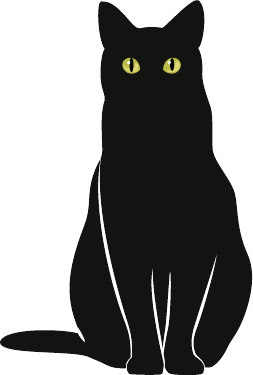}} \,\right\rangle 
    & \equiv
    \begin{pmatrix}
        0 \\
        \\
        1
    \end{pmatrix}
    \hspace*{0.5cm}
    \rightarrow \,\,\, \text{Cat awake}
    \\
    \left|\, \raisebox{-0.4em}{\includegraphics[height=1.5em]{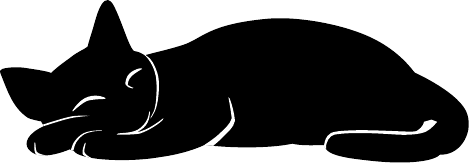}} \,\right\rangle 
    & \equiv
    \begin{pmatrix}
        1 \\
        \\
        0
    \end{pmatrix}
    \hspace*{0.5cm}
    \rightarrow \,\,\,  \text{Cat sleeping}
\end{align}
Therefore, Eq. \eqref{estado2} can be written as
\begin{align}
    \lvert \psi \rangle = 
    \alpha
    \left|\, \raisebox{-1.0em}{\includegraphics[height=2.5em]{img/gato_vivo.pdf}} \,\right\rangle
    +
    \beta
    \left|\, \raisebox{-0.5em}{\includegraphics[height=1.4em]{img/gato_morto.pdf}} \,\right\rangle
    \label{eq:estado_cat_nonormalized}
\end{align}

The quantum state $\lvert\psi\rangle$ contains all the information about all the system. Consequently, from it we must be able to compute the probability associated with each outcome of the measurement whether the cat is live or sleeping. In the Schr{\"o}dinger's cat example, the probabilities are real numbers (such as the values $9/10$ and $1/10$ in Eq. \eqref{eq:estado_cat_nonormalized}). We have then to ensure that the total probability associated with all possible outcomes of the measurement is equal to $1$.

Since $\lvert\psi\rangle$ describes the superposition of all possible states (in this case, live or sleeping), the sum of the probabilities of all these states must necessarily be equal to $1$. Mathematically, this condition is ensured by normalizing the state $\lvert\psi\rangle$, imposing
\begin{equation*}
    \langle \psi_{\text{norm}} \mid \psi_{\text{norm}} \rangle = 1,
\end{equation*}
thereby guaranteeing that the sum of the squared moduli of the coefficients — possibly complex — is exactly equal to $1$.

In other words, the coefficients appearing in the state representation are complex amplitudes, not the probabilities themselves. The probabilities are obtained from the square of the absolute value of these amplitudes:
\begin{align*}
    P_{\text{awake}} &= |\alpha|^{2},\\
    P_{\text{sleeping}} &= |\beta|^{2}.
\end{align*}

In the particular case we are looking at, the coefficient 
$\alpha$ associated with the ``awake" state is
\begin{equation*}
    \alpha = \sqrt{\frac{9}{10}} = \frac{3}{\sqrt{10}}
\end{equation*}
while for the ``sleeping" state one obtains
\begin{equation*}
    \beta = \sqrt{\frac{1}{10}} \, e^{i \theta}= \frac{1}{\sqrt{10}} \, e^{i \theta},
\end{equation*}
where $\theta$ is an arbitrary real number. Thus, the correct representation of the normalized quantum state is
\begin{equation}
    \lvert \psi_{\text{norm}}  \rangle
    = 
    \frac{3}{\sqrt{10}}
    \left|\, \raisebox{-1.0em}{\includegraphics[height=2.5em]{img/gato_vivo.pdf}} \,\right\rangle
    +
    \frac{1}{\sqrt{10}} e^{i \theta}
    \left|\, \raisebox{-0.5em}{\includegraphics[height=1.4em]{img/gato_morto.pdf}} \,\right\rangle,
    \label{eq:estado_normalizado}
\end{equation}
which is equivalent to:
\begin{equation*}
    \lvert \psi_{\text{norm}}  \rangle
    = \begin{pmatrix}
        \frac{1}{\sqrt{10}} \\
        \\
        \frac{3}{\sqrt{10}} e^{i \theta}
    \end{pmatrix}.
\end{equation*}
Note that now, when we compute
\begin{equation*}
    \langle \psi_{\text{norm}} \lvert \psi_{\text{norm}} \rangle
    =
    \left(\frac{3}{\sqrt{10}}\right)^2
    +
    \left(\frac{1}{\sqrt{10}}\right)^2
    =
    1.
\end{equation*}
From now on, we will omit the subscript $_{\text{norm}}$. Using this notation, in order to compute the probability of the state $\lvert i \rangle$ (where the index $i$ may refer to the live or sleeping states), it is then sufficient to evaluate 
\begin{equation}
    P_i = \lvert \langle i \mid \psi \rangle \rvert^{2},
    \label{eq:probability}
\end{equation}
that is, for the state \eqref{eq:estado_normalizado}:
\begin{align*}
    \left\langle \, \raisebox{-0.5em}{\includegraphics[height=1.2em]{img/gato_vivo.pdf}} \, \right| \psi \Big\rangle &= \begin{pmatrix}
        0 \ \ \ 1
    \end{pmatrix}\begin{pmatrix}
        \frac{1}{\sqrt{10}} \\
          \\
        \frac{3}{\sqrt{10}}
    \end{pmatrix} = \frac{3}{\sqrt{10}} \ \ \ \ \ \therefore \ \ \ \ \  P_{awake} = \left| \left\langle \, \raisebox{-0.5em}{\includegraphics[height=1.7em]{img/gato_vivo.pdf}} \, \right| \psi \Big\rangle \right|^2 = \frac{9}{10} \\
    \Big\langle \, \raisebox{-0.35em}{\includegraphics[height=1.0em]{img/gato_morto.pdf}} \, \Big| \psi \Big\rangle &= \begin{pmatrix}
        1 \ \ \ 0
    \end{pmatrix}\begin{pmatrix}
        \frac{1}{\sqrt{10}} \\
          \\
        \frac{3}{\sqrt{10}}
    \end{pmatrix} = \frac{1}{\sqrt{10}} \ \ \ \ \ \therefore \ \ \ \ \  P_{sleeping} = \left| \Big\langle \, \raisebox{-0.35em}{\includegraphics[height=1.0em]{img/gato_morto.pdf}} \, \Big| \psi \Big\rangle \right|^2 = \frac{1}{10}
\end{align*}

Crucially, the vector $|\psi\rangle$ can be expressed as a linear combination (or superposition) of different basis states $|i\rangle$, which correspond to the possible outcomes of a measurement. This can be then written as
\begin{equation}
    \lvert\psi\rangle = \sum_i C_i |i\rangle,
\end{equation}
where $C_i$ are complex numbers called probability amplitudes. The square of the absolute value of these amplitudes, $\lvert C_i \lvert^2$, gives the probability of obtaining the state $|i\rangle$ when performing a measurement of an observable having as eigenvalues exactly the states $\{|i\rangle\}$.

\subsection{The energy levels}

Particles such as electrons in an atom, when they have a negative energy and are in a bound state, cannot occupy just any energy level, but are restricted to discrete, or quantized, values. These particles occupy only certain allowed states, each with a well-defined energy value \citep{Cohen-Tannoudji:101367,Shankar,griffiths2018,Sakurai_Napolitano_2020}.

An extremely useful model in atomic physics is the so-called {\it two-level system}. Although atoms have of course many possible energy levels \citep{Foot2005}, in many situations we can simplify their description 
using only two energy levels \citep{Orszag2024}: in other words, when only (or mostly) contributions from two energy levels enter the physical quantities, we can restrict the basis to the two energy levels
\begin{itemize}
    \item a state $\lvert \estadog \rangle$ with energy $E_{\estadog}$;
    \item a state $\lvert \estadoe \rangle$ with energy $E_{\estadoe}$.
\end{itemize}
$\lvert\estadog\rangle$ and $\lvert\estadoe\rangle$ are eigenstates of the Hamiltonian $H$:
\begin{align}
    H \lvert \estadog \rangle &= E_{\estadog}\lvert \estadog \rangle, \\
    H \lvert \estadoe \rangle &= E_{\estadoe}\lvert \estadoe \rangle.
\label{eig}
\end{align}
We refer to $\lvert \estadog \rangle$ as the {\it ground state} and to $\lvert \estadoe \rangle$ as the {\it excited state}. As an example, for a particle in a double well potential, having two energy minima well separated by an energy barrier, the state $\lvert \estadog \rangle$ and $\lvert \estadog \rangle$ can be really the two lowest energy states, and the $\lvert \estadog \rangle$ the state with minimum energy.

The energy difference between the two energies $E_{\estadoe}$ and $E_{\estadog}$ defines the frequency of transition 
$$\omega = \frac{E_{\estadoe} - E_{\estadog}}{\hbar}$$
in the sense that a 
photon can be absorbed, if its angular frequency  $\omega$, is such that the atom may change its state from $|\estadog\rangle$ to $|\estadoe\rangle$. Inversely, when the atom state decays spontaneously from $|\estadoe\rangle$ to $|\estadog\rangle$, it emits a photon of the same energy.

An important point to be noticed is that all the parameters of the Hamiltonian of the system, such as the mass, are encoded in the energies $E_{\estadog}, E_{\estadoe}$. If one wants to study the motion of the atoms, one has to include the kinetic energy term explicitly (see Section \ref{sec:threeLevel}).

The average energy of the atom can be calculated as the average of the energies of the levels that constitute the system ($E_{\estadog}$ and $E_{\estadoe}$) weighted by the probabilities of each of these states ($P_{\estadog}$ and $P_{\estadoe}$), that is
\begin{align*}
    \langle E \rangle &= E_b \, P_\estadoe \, + \, E_a \, P_\estadog. 
\end{align*}
By using Eq. \eqref{eq:probability}, we can rewrite the average energy as 
\begin{align*}
    \langle E \rangle &= E_\estadoe \, \lvert \langle\estadoe\lvert\psi\rangle \lvert^2 \, +  \, E_\estadog \, \lvert \langle\estadog\lvert\psi\rangle \lvert^2 \\
     &= \langle \psi \lvert \left(E_\estadoe \lvert \estadoe \rangle \langle \estadoe \lvert \, + \, E_\estadog \lvert \estadog \rangle \langle \estadog \lvert \right) \lvert \psi \rangle
\end{align*}
Based on the previous expression, we can define the Hamiltonian operator in the basis $\langle\estadog\lvert, \langle\estadoe\lvert$ as:
\begin{equation*}
    H = E_\estadoe \lvert \estadoe \rangle \langle \estadoe \lvert \, + \, E_\estadog \lvert \estadog \rangle \langle \estadog \lvert
\end{equation*}
which in turn can be rewritten in matrix form, see e.g. \citep{kok2018first}, as
\begin{equation*}
    H = \begin{pmatrix}
        E_\estadoe & 0 \\
        0 & E_\estadog
    \end{pmatrix}.
    \label{ham:matrix}
\end{equation*}

\subsection{The Schr{\"o}dinger equation}

The Hamiltonian not only encodes the energy of the system, but it is also related to the time evolution of the system. Thus, once the Hamiltonian of a quantum system is known, it becomes possible to determine how its state evolves over time \citep{Cohen-Tannoudji:101367,Shankar, griffiths2018,kok2018first,Sakurai_Napolitano_2020}. Consider again the state of a two-level atom at time ($t = 0$). We write the initial state as a coherent superposition of the energy eigenstates as
\begin{equation*}
    \lvert \psi(0) \rangle = C_a \lvert \estadog \rangle + C_b \lvert \estadoe \rangle ,
\end{equation*}
where $C_a$ e $C_b$ are complex amplitudes satisfying  $|C_a|^2 + |C_b|^2 = 1$.

Since $\lvert\estadog\rangle$ and $\lvert\estadoe\rangle$ are eigenstates of the Hamiltonian according \eqref{eig}, the time evolution of the probability amplitudes $C_a, C_b$ is simply a phase factor. The state at time $t$ becomes
\begin{equation*}
    \lvert \psi(t) \rangle = C_a e^{-iE_{\estadog}t/\hbar} \lvert \estadog \rangle + C_b e^{-iE_{\estadoe}t/\hbar} \lvert \estadoe \rangle.
\end{equation*}
This expression shows that each component of the state acquires a phase proportional to its energy. Importantly, the relative phase between the levels evolves as
 \begin{equation*}
    \Delta \phi(t) = \frac{E_{\estadoe}-E_{\estadog}}{\hbar} \, t = \omega t,
\end{equation*}
where $\omega$ is the transition frequency of the two-level system. Computing the time derivative of the state and 
using the eigenvalue equations of the Hamiltonian \eqref{eig}, we obtain the celebrated time-dependent Schr{\"o}dinger equation:
\begin{equation}
     i\hbar \frac{d}{dt} \lvert \psi(t) \rangle = H \lvert \psi(t) \rangle .
\end{equation}

This is the central equation governing the quantum evolution of non-relativistic quantum systems. It tells us that once the Hamiltonian is known, the time dynamics of the system is completely determined. For the two-level atom with Hamiltonian \eqref{ham:matrix}, 
the Schr{\"o}dinger equation describes how population and phase evolve between the ground and excited states.
In the next sections, we will incorporate interactions such as laser fields, which modify the Hamiltonian and allow transitions between the levels, leading to the quantum dynamics fundamental to atomic interferometry and quantum sensing.

\section{The two-level system}\label{section:2levels}

\begin{figure}[H]
	\centering
    \includegraphics[width=1.\textwidth]{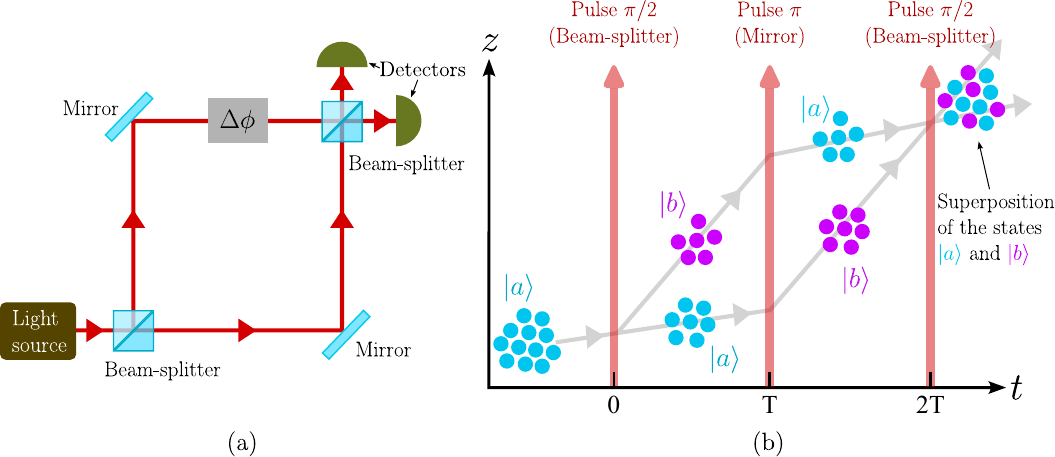}
	\caption{(a) Schematic representation of a 
    Mach–Zehnder interferometer based on optical beams. 
(b) Illustration of a quantum Mach–Zehnder interferometer, in which matter waves are used.}
	\label{fig:equivalenciaInter}
\end{figure}

Quantum gravimeters use the principle of interferometry to perform precise measurements of gravitational acceleration. More specifically, they operate as Mach-Zehnder interferometers that utilize the phenomenon of interference to measure the phase differences between two distinct paths 
of a wave \citep{Freier_2016, Schilling_2020, JM2020}. In a typical realization of a Mach-Zehnder device, a light source emits a beam that initially 
enters a so-called {\it beam splitter}, causing the wave to propagate through two different arms (paths). Each beam is then reflected, e.g. by mirrors, until they are recombined and, subsequently, split again by another beam splitter. Finally, two light intensity detectors measure the interference pattern generated by the recombination of the light beams, which is a function of the phase difference $\Delta\phi$ between the two light beams that traverse distinct paths (Figure \ref{fig:equivalenciaInter}.a).

A quantum Mach-Zehnder interferometer
may probe also the interference of matter waves. In this case, there are no physical beam splitters or mirrors as for electromagnetic waves. Laser pulses act on atoms in a manner equivalent to a beam splitter or a mirror, depending on how the laser-atom interaction takes place. What determines whether the laser acts as a beam splitter or as a mirror is the relation between the frequencies involved and the duration of the pulse. This aspect will be discussed in more detail throughout this section. For the purposes of this work, a $\pi/2$ pulse will be defined as the pulse that acts as a beam splitter, while a $\pi$ pulse will be defined as the pulse that acts in a manner equivalent to a mirror.

In a quantum Mach-Zehnder interferometer, the atoms can be initially prepared in the same internal state $|a\rangle$ (the ground state) and, after a $\pi/2$ pulse, are ``split'' into two paths, which are described by atoms in the different internal states $|a\rangle$ and $|b\rangle$. Atoms in the state $|a\rangle$ follow a trajectory that is different from that followed by atoms in the state $|b\rangle$ (excited state). The atoms then propagate along two distinct paths of the interferometer and, after a time interval $T$, interact with a $\pi$ pulse, which induces an inversion of the quantum states. Subsequently, after a further time interval $T$, the paths are recombined, and a third interaction with a $\pi/2$ laser pulse takes place, resulting in a cloud of atoms in a superposition of the states $|a\rangle$ and $|b\rangle$ (Figure \ref{fig:equivalenciaInter}.b).

To make contact with the the notions reminded in the previous section, a quantum gravimeter can be conceptually simplified as a two-level system, with two states: the ground state ($\lvert \estadog \rangle$) and the excited state ($\lvert \estadoe \rangle$). It is important to note that this system exhibits two levels in a manner analogous to Schr\"{o}dinger's cat thought experiment. Thus, by comparison and only for didactic purposes, one may associate the level $\lvert \estadog \rangle$ with the atom being in the ``awake" state, whereas $\lvert \estadoe \rangle$ corresponds to what is the ``sleeping" state.

In this context, the following question arises: what is the analogue of the ``radiation detector" in the quantum gravimeter? In other words, what is the mechanism responsible for inducing transitions between the atomic levels in the quantum gravimeter? The answer is: a laser pulse. In the quantum gravimeter the particle is initially in the ground state, characterized by the energy $E_{a}$. After interacting with a laser pulse, the particle may absorb a photon and, consequently, undergo a transition to the excited state, with energy $E_{b}$.
\begin{figure}[H]
	\centering
    \includegraphics[width=.8\textwidth]{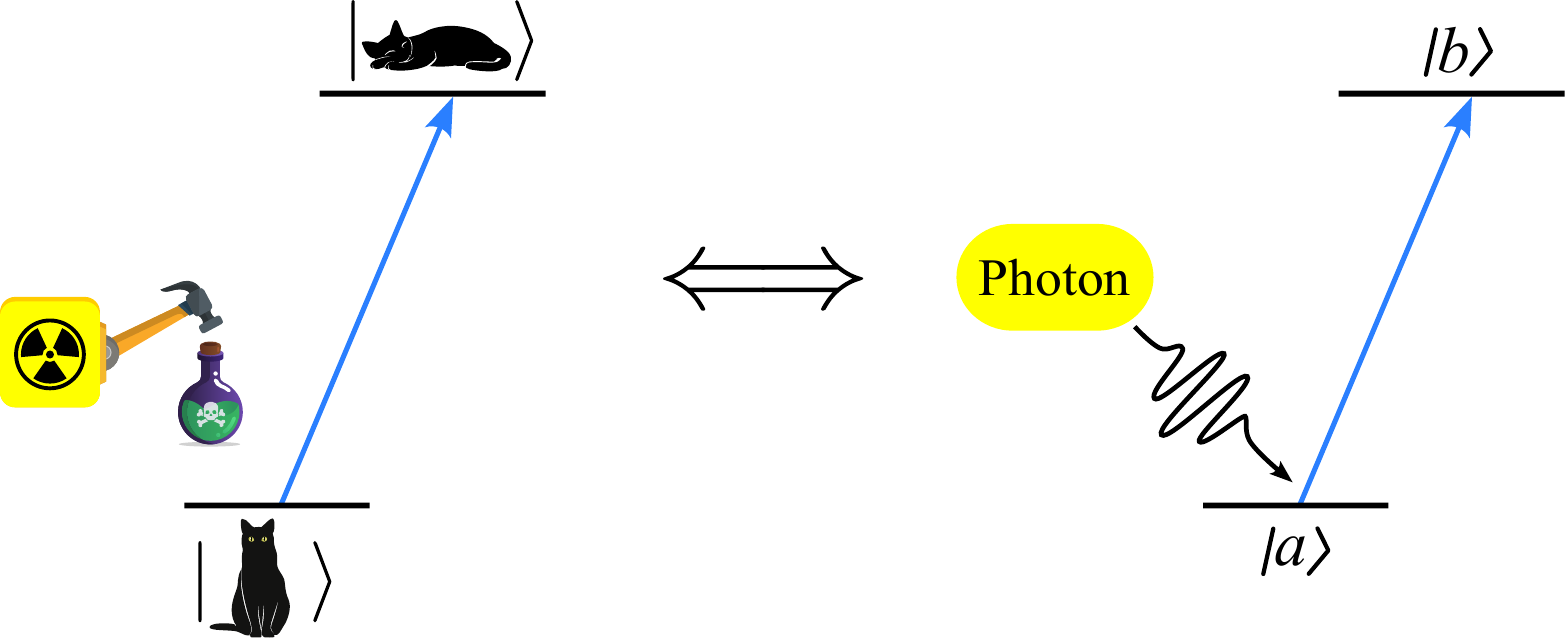}
	\caption{A pictorial comparison between the operating principles of a quantum gravimeter and the Schr{\"o}dinger's cat thought experiment. The figure illustrates that the ``alive" (or "awake") and ``dead" (or "sleeping") states of the cat are conceptually analogous to the atomic energy levels. In this analogy, the laser photons play a role equivalent to that of the radiation detector in the cat experiment. Specifically, the absorption of a photon by the atom determines whether a transition between energy levels will occur, in the same way that poison/sleeping substance is released (or not) 
    in the Schr{\"o}dinger's cat thought experiment.}
	\label{fig:equivalencia}
\end{figure}

\begin{figure}[H]
	\centering
    \includegraphics[width=.25\textwidth]{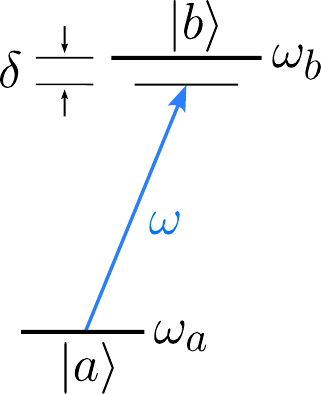}
	\caption{Schematic representation of the two levels with their associated frequencies.}
	\label{fig:2levels}
\end{figure}

For the state transition to occur, the laser frequency $\omega$ must be resonant with the energy difference between the two levels, so that the photon provides the necessary energy to induce the transition \citep{Orszag2024}. Thus, the resonance condition requires that $\omega \approx \omega_{ba}$, where $\omega_{ba} = \omega_b - \omega_a$ is the frequency associated with the transition \citep{Orszag2024, YOUNG1997}. The difference between the laser frequency and the transition frequency defines the {\it detuning}, given by $$\delta = \omega - \omega_{ba}=\omega-(\omega_b-\omega_a)$$ (see figure \ref{fig:2levels}).

The electric field of the laser can be written as:
\begin{equation}
	\mathbf{E} = \mathbf{E}_0 \, \cos\left(\omega t + \phi\right).
    \label{eq:campoEsimples}
\end{equation}
Since the dimension of atoms is on the order of angstroms ($\approx 10^{-10}\,\mathrm{m}$), while the wavelength of $\mathbf{E}$ is typically much larger, 
the amplitude of the field is essentially constant over the spatial extension of the atom. Consequently, the atom may be treated as a dipole of moment $\mathbf{d}$ under the influence of the electromagnetic field, and therefore the Hamiltonian describing the system can be written in matrix form as:
\begin{equation}
    \hat{H}_{\text{int}}  = -\mathbf{d}\cdot\mathbf{E}= \begin{pmatrix}
		0 & & \langle \estadoe \lvert -\mathbf{d}\cdot\mathbf{E}\lvert \estadog \rangle \\
		& & \\
        \langle \estadog \lvert -\mathbf{d}\cdot\mathbf{E}\lvert \estadoe \rangle & & 0
	\end{pmatrix},
\end{equation}
Therefore the Hamiltonian of an atom interacting with an electromagnetic field can be written as the sum of two contributions:
\begin{equation*}
    \hat{H} = \hat{H}_0 + \hat{H}_{\text{int}},
\end{equation*}
where $\hat{H}_0$ corresponds to the time-independent Hamiltonian of the free atom given by
\begin{align}
	\hat{H}_{0} = \hbar \wg \projg \, + \, \hbar \we \proje \, = \begin{pmatrix}
		\hbar \we & & 0\\
		& & \\
        0 & & \hbar \wg
	\end{pmatrix},
    \label{eq:H02niveis}
\end{align}
with $(0 \ 1)^T=|a\rangle$ and $(1 \ 0)^T=|b\rangle$ (of course, this choice is arbitrary and we could have choosen the opposite). The Hamiltonian $\hat{H}$ can be written as follows:
\begin{equation}
	\hat{H} = \hbar \wg \projg \, + \, \hbar \we \proje \, - \,
	 \mathbf{d}\cdot\mathbf{E}
\end{equation}
The time evolution of a general quantum state is described by
\begin{equation}
	\lvert \psi(t) \rangle = \Cg \, e^{-i \wg t} \ketg + \Ce \, e^{-i \we t} \kete,
    \label{eq:psi2level}
\end{equation}
with $\Cg$ and $\Ce$ complex amplitudes. This evolution is governed by the time-dependent 
Schr\"{o}dinger equation:
\begin{equation}
	i \hbar \frac{d}{dt}\lvert \psi(t) \rangle = \left(\hat{H}_0 + \hat{H}_{\text{int}}\right) \lvert \psi(t) \rangle
    \label{eq:equacaoschrondigertotal}
\end{equation}
By differentiating \eqref{eq:psi2level} with respect to time, we obtain
\begin{align*}
	i\hbar\frac{\partial}{\partial t}\lvert \psi(t) \rangle =& \, i\hbar\dot{C}_\estadog \, e^{-i \wg t} \ketg + \hbar \wg \Cg \, e^{-i \wg t} \ketg + \\
    & \, + i\hbar\dot{C}_\estadoe \, e^{-i \we t} \kete + \hbar \we \Ce \, e^{-i \we t} \kete \\ 
    & \, = i\hbar\dot{C}_\estadog \, e^{-i \wg t} \ketg + i\hbar\dot{C}_\estadoe \, e^{-i \we t} \kete + \hat{H}_0 \lvert \psi(t) \rangle,
\end{align*}
with
\begin{align*}
	\hat{H}_{\text{int}} \lvert \psi(t) \rangle =& \, i\hbar\dot{C}_\estadog \, e^{-i \wg t} \ketg + i\hbar\dot{C}_\estadoe \, e^{-i \we t} \kete.
\end{align*}
To proceed, we project this expression onto the basis states $\ketg$ and $\kete$. This yields
\begin{align*}
    \langle \estadog\lvert\hat{H}_{\text{int}} \lvert \psi(t) \rangle =& i\hbar\dot{C}_\estadog \, e^{-i \wg t}, \\
    \langle \estadoe\lvert\hat{H}_{\text{int}} \lvert \psi(t) \rangle =& i\hbar\dot{C}_\estadoe \, e^{-i \we t}.
\end{align*}
Thus,
\begin{align}
    \langle \estadog\lvert\hat{H}_{\text{int}} \lvert \psi(t) \rangle =& \, \langle \estadog\lvert -\mathbf{d}\cdot\mathbf{E} \lvert \estadoe \rangle \Ce e^{-i\we t}, \label{eq:aHpsi0} \\
    \langle \estadoe\lvert\hat{H}_{\text{int}} \lvert \psi(t) \rangle =& \, \langle \estadoe\lvert -\mathbf{d}\cdot\mathbf{E} \lvert \estadog \rangle \Cg e^{-i\wg t}. \label{eq:bHpsi0}
\end{align}

The electric field $\mathbf{E}$, given in Eq. \eqref{eq:campoEsimples} and appearing in the expressions, above can be decomposed as:
\begin{equation}
	\mathbf{E} = \mathbf{E}_0 \, \cos\left(\omega t + \phi\right) = \mathbf{E}_0 \, \left( \frac{e^{i\omega t}e^{i\phi} \, + \, e^{-i\omega t}e^{-i\phi}}{2} \right).
\end{equation}
This decomposition allows the application of the so-called Rotating Wave Approximation (RWA) in order to neglect the rapidly oscillating term \citep{Orszag2024} and to retain only the slowly varying components of the field. This means that only one of the complex exponentials, either $e^{i\omega t}$ or $e^{-i\omega t}$, contributes significantly to the interaction, while the other may be discarded because it oscillates much faster than the system dynamics. In the case of the transition $\lvert g\rangle \rightarrow \lvert e\rangle$, the rapidly oscillating component $e^{i\omega t}$ can be eliminated, and only $e^{-i\omega t}$ is retained, as it is responsible for photon absorption. Conversely, for the transition $\lvert e\rangle \rightarrow \lvert g\rangle$, the term to be kept is $e^{i\omega t}$, corresponding to the stimulated emission process.

With this simplification, Eqs. \eqref{eq:aHpsi0} and \eqref{eq:bHpsi0} take the form
\begin{align*}
    \langle \estadog\lvert -\mathbf{d}\cdot\mathbf{E} \lvert \estadoe \rangle \Ce e^{-i\we t} =& \frac{\langle \estadog\lvert -\mathbf{d}\cdot\mathbf{E}_0 \lvert \estadoe \rangle}{2} \, \Ce e^{-i\we t} e^{i\omega t}e^{i\phi}, \\
    \langle \estadoe\lvert -\mathbf{d}\cdot\mathbf{E} \lvert \estadog \rangle \Cg e^{-i\wg t}  =& \frac{\langle \estadoe\lvert -\mathbf{d}\cdot\mathbf{E}_0 \lvert \estadog \rangle}{2} \, \Cg e^{-i\wg t} e^{-i\omega t}e^{-i\phi}.
\end{align*}

Rewriting these equations, we obtain
\begin{align*}
    i\dot{C}_\estadog =& \frac{\langle \estadog\lvert -\mathbf{d}\cdot\mathbf{E}_0 \lvert \estadoe \rangle}{2\hbar} \, \Ce e^{i[\omega - (\we-\wg)] t} e^{i\phi} \\
    i\dot{C}_\estadoe =& \frac{\langle \estadoe\lvert -\mathbf{d}\cdot\mathbf{E}_0 \lvert \estadog \rangle}{2\hbar} \,  \Cg e^{-i[\omega - (\we-\wg)] t} e^{-i\phi}.
\end{align*}

Introducing the Rabi frequency \citep{Scully_Zubairy_1997}
\begin{equation*}
    \Omega_{\estadoe\estadog} = \frac{\langle \estadoe\lvert -\mathbf{d}\cdot\mathbf{E}_0 \lvert \estadog \rangle}{\hbar},
\end{equation*}
one has 
\begin{equation} 
    \Omega_{\estadoe\estadog}^{*} = \Omega_{\estadog\estadoe} = \frac{\langle \estadog\lvert -\mathbf{d}\cdot\mathbf{E}_0 \lvert \estadoe \rangle}{\hbar}.
\end{equation}
Then, the system of coupled equations acquires the form:
\begin{align*}
	i \, \dot{C}_\estadog =& \frac{\Omega_{\estadoe\estadog}^{*}}{2} \, e^{+i\left(\delta\,t + \phi\right)} \, C_{\estadoe}, \\
    i \, \dot{C}_\estadoe =& \frac{\Omega_{\estadoe\estadog}}{2} \, e^{-i\left(\delta\,t + \phi\right)} \, C_{\estadog}.
\end{align*}
This allows us to define $\hat{H}_{\text{int}}$ as
\begin{align}
	\hat{H}_{\text{int}} = \frac{\hbar}{2}\begin{pmatrix}
		0 & \Omega_{\estadoe\estadog} \, e^{-i\left(\delta\,t + \phi\right)}\\
		&  \\
		\Omega_{\estadoe\estadog}^{*} \, e^{+i\left(\delta\,t + \phi\right)} & 0
	\end{pmatrix}
    \label{eq:HamIntp2Level}
\end{align}

\begin{figure}[H]
	\centering
	\includegraphics[width=0.25\paperwidth]{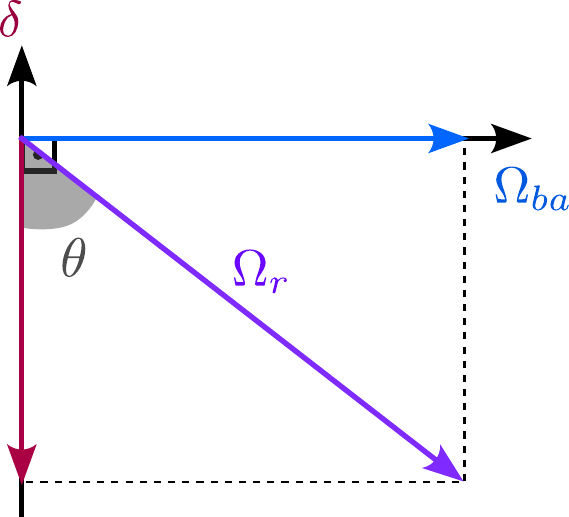}
    \caption{Geometric interpretation of the relationship between $\theta$ and the parameters $\delta$, $\Omega_{ba}$ and $\Omega_{R}$.}
    \label{fig:esquema}
\end{figure}

The  Hamiltonian can be made time-independent by adopting a rotating reference frame, expressing the physical idea that during a short pulse can be considered approximately constant.  
This is achieved by transforming the state ($\lvert \psi \rangle$) into the state in the rotating frame, ($\lvert \psi \rangle_R$), defined by a rotation around the ($\boldsymbol{\hat{z}}$) axis with angle ($-\delta t$):
\begin{equation*}
	\lvert \psi \rangle_R = \boldsymbol{D} \left(\boldsymbol{\hat{z}}, \, -\delta t\right)\lvert \psi \rangle
\end{equation*}
where $\boldsymbol{D}$ is defined as \citep{YOUNG1997}:
\begin{align}
	\boldsymbol{D} = \begin{pmatrix}
		e^{+i\delta\,t/2} & 0 \\
		0 & e^{-i\delta\,t/2}
	\end{pmatrix}. \label{eq:Drot}
\end{align}
Applying the Schr\"{o}dinger equation to the state expressed in the rotating frame yields:
\begin{align}
	i \hbar \frac{d}{dt}\left(\boldsymbol{D}^{\dagger}\lvert \psi \rangle_R\right) &= \hat{H}_{\text{int}}  \left(\boldsymbol{D}^{\dagger}\lvert \psi \rangle_R\right)
\end{align}
Rearranging the terms gives the time evolution equation for the state in the rotating frame:
\begin{align}
	i \hbar \frac{d \lvert \psi \rangle_R}{dt} &= \left(\boldsymbol{D} \hat{H}_{\text{int}} \boldsymbol{D}^{\dagger} - i \hbar \boldsymbol{D} \frac{d \boldsymbol{D}^{\dagger}}{dt} \right) \, \lvert \psi\rangle_R . \label{eq:rotatingH_R}
\end{align}
Therefore, the Hamiltonian in the rotating reference frame is time-independent and it is given by:
\begin{equation}
	\hat{H}_R = \boldsymbol{D} \hat{H}_{\text{int}} \boldsymbol{D}^{\dagger} - i \hbar \boldsymbol{D} \frac{d \boldsymbol{D}^{\dagger}}{dt}= \frac{\hbar}{2}\begin{pmatrix}
		-\delta & \Omega_{\estadoe\estadog} \, e^{-i\phi}\\
		& \\
		\Omega_{\estadoe\estadog}^{*} \, e^{+i\phi} & +\delta
	\end{pmatrix}.
    \label{eq:H_Rot}
\end{equation}
The eigenvalues are 
\begin{equation}
\lambda_{\pm} = \pm \frac{\hbar \Omega_R}{2}, 
\label{eq:sigvalues}
\end{equation}
where $\Omega_R$ is the so-called (off-resonant) Rabi frequency given by $$\Omega_R = \sqrt{\Omega_{\estadoe\estadog}^2 + \delta^2}$$
(from now on, in the main text and in Figure
\ref{fig:esquema}, unless otherwise specified, we denote the modulus of 
$\Omega_{\estadoe\estadog}$ simply by $\Omega_{\estadoe\estadog}$). 
Accordingly, we can express $\delta$, 
$\Omega_{\estadoe\estadog}$, and $\Omega_R$ in terms of an  angle $\theta$, defined as follows (see Figure \ref{fig:esquema}):
\begin{align}
    \sin \theta \, = \, \frac{\Omega_{\estadoe\estadog}}{\Omega_R}  \hspace{2cm} \cos \theta \, = \, -\frac{\delta}{\Omega_R}
    \label{eq:sincos}
\end{align}
Note that, when $\delta \ll \Omega_{\estadoe\estadog}$ we have $\Omega_{\estadoe\estadog} \approx \Omega_R$. Therefore, under the resonant condition ($\delta = 0$), the angle $\theta$ becomes $\theta = \pi/2$, so that 
$\sin \theta  =  1$ and  $\cos \theta = 0$.

The Hamiltonian $\hat{H}_R$ is diagonalized in terms of the eigenvectors $|\lambda_{+}\rangle_R$ and $|\lambda_{-}\rangle_R$, which are associated with the eigenvalues $\lambda_{+}$ and $\lambda_{-}$, respectively (Appendix \ref{secA1}). The eigenstates expressed in the rotating-frame basis $\{\lvert \estadog\rangle_R, \lvert \estadoe\rangle_R\}$, where again $(0 \ 1)^T=\lvert \estadog\rangle_R$ and $(1 \ 0)^T=\lvert \estadoe\rangle_R$, are given by
\begin{align}
    \lvert \lambda_{+} \rangle &= \sin\left(\frac{\theta}{2}\right) e^{+i\frac{\phi}{2}} \lvert \estadog \rangle_R + \cos\left(\frac{\theta}{2}\right) e^{-i\frac{\phi}{2}} \lvert \estadoe \rangle_R \nonumber \\
    \lvert \lambda_{-} \rangle &= \cos\left(\frac{\theta}{2}\right) e^{+i\frac{\phi}{2}} \lvert \estadog \rangle_R -\sin\left(\frac{\theta}{2}\right) e^{-i\frac{\phi}{2}} \lvert \estadoe \rangle_R
\end{align}
The projectors $\lvert \lambda_{+}\rangle\langle\lambda_{+}\lvert$ and $\lvert\lambda_{-}\rangle\langle\lambda_{-}\lvert$ take the matrix forms given in Eqs. \eqref{eq:projlambp}--\eqref{eq:projlambm}.  
Using Eqs. \eqref{eq:Drot}, \eqref{eq:sigvalues}] and \eqref{eq:psirot} 
gives:
\begin{align}
	\lvert \lambda_{+}\rangle\langle\lambda_{+}\lvert &= \begin{pmatrix}
		\cos^2\left(\frac{\theta}{2}\right) & \frac{\sin \theta}{2} e^{-i\phi} \\
         & \\
		\frac{\sin \theta}{2} e^{+i\phi} & \sin^2\left(\frac{\theta}{2}\right)
	\end{pmatrix} \label{eq:projlambp}
\end{align}
\begin{align}
    \lvert \lambda_{-}\rangle\langle\lambda_{-}\lvert &= \begin{pmatrix}
		\sin^2\left(\frac{\theta}{2}\right) & -\frac{\sin \theta}{2} e^{-i\phi} \\
         & \\
		-\frac{\sin \theta}{2} e^{+i\phi} & \cos^2\left(\frac{\theta}{2}\right)
	\end{pmatrix} \label{eq:projlambm}
\end{align}

The time evolution of the atomic state, in the rotating frame and during the interaction with a light pulse starting at time $t=t_0$, is given by
\begin{equation}
	{\lvert \psi \left(t_0 \, + \, \tau\right) \rangle}_{R} = \left( e^{-\frac{i\,\lambda_{+}\tau}{\hbar}}  \lvert \lambda_{+}\rangle\langle\lambda_{+}\lvert \ + \ e^{-\frac{i\,\lambda_{-}\tau}{\hbar}}  \lvert \lambda_{-}\rangle\langle\lambda_{-}\lvert\right) \boldsymbol{D} \left(\boldsymbol{\hat{z}}, \, -\delta t\right)\lvert \psi\left(t_0 \right) \rangle. \label{eq:psirot}
\end{equation}
It follows
\begin{align}
	{\lvert \psi \left(t_0  +  \tau\right) \rangle} = \begin{pmatrix}
		e^{-i\delta\tau/2} \left[ \cos\left(\frac{\Omega_R\tau}{2}\right) - i\cos\theta\,\sin\left( \frac{\Omega_R\tau}{2} \right) \right] & -i e^{-i\delta\tau/2} e^{-i(\delta t_0 + \phi)} \sin\theta\sin\left(\frac{\Omega_R\tau}{2}\right) \\
         & \\
         & \\
		-i e^{i\delta\tau/2} e^{i(\delta t_0 + \phi)} \sin\theta\sin\left(\frac{\Omega_R\tau}{2}\right) & e^{i\delta\tau/2} \left[ \cos\left(\frac{\Omega_R\tau}{2}\right) + i\cos\theta\,\sin\left( \frac{\Omega_R\tau}{2} \right) \right]
	\end{pmatrix} \times \nonumber \\
    \times \begin{pmatrix}
		C_{\estadoe}(t_0)\\
         & \\
		C_{\estadog}(t_0).
	\end{pmatrix}
    \label{eq:psit0tau}
\end{align}
This gives the final result:
\begin{align}
    C_{\estadoe}(t_0+\tau) = e^{-i\delta\tau/2}\bigg\{  &C_{\estadoe}(t_0) \left[ \cos\left(\frac{\Omega_R\tau}{2}\right) - i\cos\theta\,\sin\left( \frac{\Omega_R\tau}{2} \right) \right] + \nonumber\\
    &+ C_{\estadog}(t_0) e^{-i(\delta t_0 + \phi)} \sin\theta\sin\left(\frac{\Omega_R\tau}{2}\right) \bigg\},\\
	C_{\estadog}(t_0+\tau) = e^{+i\delta\tau/2}\bigg\{  &C_{\estadoe}(t_0) \left[ -i e^{i(\delta t_0 + \phi)} \sin\theta\sin\left(\frac{\Omega_R\tau}{2}\right) \right] + \nonumber \\
    &+C_{\estadog}(t_0) \left[ \cos\left(\frac{\Omega_R\tau}{2}\right) + i\cos\theta\,\sin\left( \frac{\Omega_R\tau}{2} \right) \right] \bigg\}.
\end{align}

For $|\delta| \ll \Omega_R$ one has
\begin{align}
    C_{\estadoe}(t_0+\tau) &= e^{-i\delta\tau/2}\left[  C_{\estadoe}(t_0) \cos\left(\frac{\Omega_R\tau}{2}\right) - \, i\,  C_{\estadog} e^{-i(\delta t_0 + \phi)} \sin\left(\frac{\Omega_R\tau}{2}\right) \right];\\
	C_{\estadog}(t_0+\tau) &= e^{+i\delta\tau/2}\left[-i C_{\estadoe}(t_0) e^{i(\delta t_0 + \phi)} \sin\left(\frac{\Omega_R\tau}{2}\right) +  C_{\estadog}(t_0) \cos\left(\frac{\Omega_R\tau}{2}\right) \right].
\end{align}

For a pulse with duration $\tau = \pi/\Omega_{R}$ the evolution operator corresponds to a 
$\pi$-pulse. In this regime, the pulse performs a complete population inversion between the two internal states $\lvert \estadog \rangle$ and $\lvert \estadoe \rangle$, apart from well-defined phase factors that depend on the detuning $\delta$, the laser phase $\phi$, and the pulse start time $t_{0}$. Consequently, the amplitude evolution simplifies significantly, 
since
$\cos\left(\frac{\Omega_{R}\tau}{2}\right) = \cos\left(\frac{\pi}{2}\right) = 0$ and 
$\sin\left(\frac{\Omega_{R}\tau}{2}\right) = \sin\left(\frac{\pi}{2}\right) = 1$.
Substituting these values into the general expressions for the amplitudes yields the well-known inversion of the occupation of levels, associated with a pulse duration of $\tau$:
\begin{align}
	C_{\estadoe}(t_0+\tau) &= - i\,  C_{\estadog}(t_0) \, e^{-i(\delta t_0 + \phi)} e^{-i\delta\tau/2} \label{eq:coeficiente_ce};\\
	C_{\estadog}(t_0+\tau) &= -i \, C_{\estadoe}(t_0) \, e^{i(\delta t_0 + \phi)} e^{i\delta\tau/2} 
    \label{eq:coeficiente_cg}
\end{align}

Eqs. \ref{eq:coeficiente_ce} and \ref{eq:coeficiente_cg} provide the evolution of the coefficients $C_a$ and $C_b$. They must be applied individually to each pulse of the sequence that will be momentarily introduced, considering the $\pi/2 - \pi - \pi/2$ pulse sequence. Therefore, the time $t_0$ indicated in Eqs. \ref{eq:coeficiente_ce} and \ref{eq:coeficiente_cg} 
represents the moment when each pulse initiates. In this way, the evolution of the coefficients for each quantum state over time can be determined based on the starting times of each pulse, as described for the following pulse sequence.
\begin{figure}[H]
	\centering
    \includegraphics[width=.45\textwidth]{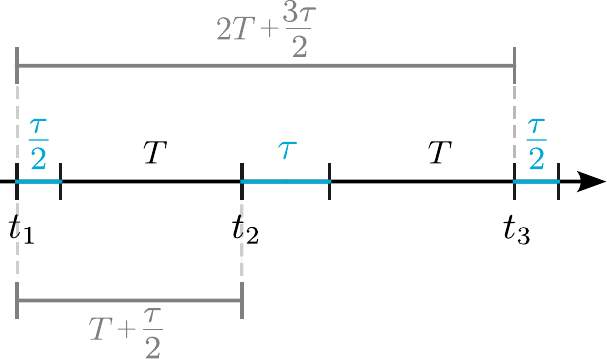}
	\caption{Pulse sequence discussed in the text. The first pulse, with a duration of $\tau/2$, starts at time $t_1$. The second pulse, with a duration of $\tau$, starts at time $t_2 = t_1 + T + \tau/2$. The third pulse starts at $t_3 = t_1 + 2T + 3\tau/2$ with a duration of $\tau/2$.}
	\label{fig:timearrow}
\end{figure}

We now define the pulse sequence, displayed in Figure \ref{fig:timearrow}. The first pulse, which has a duration of $\tau/2$ and a phase $\phi_1$, occurs at time $t_1$. The second pulse, with a duration $\tau$ and phase $\phi_2$, occurs at time $t_2 = t_1 + T + \tau/2$. Finally, the third pulse is again a $\pi/2$ pulse (with duration $\tau/2$ and phase $\phi_3$), occurring at $t_3 = t_1 + 2T + 3\tau/2$. Based on this sequence, and using Eqs. \eqref{eq:coeficiente_ce} and \eqref{eq:coeficiente_cg}, together with the initial conditions $C_{\estadog}(t_1) = 1$ and $C_{\estadoe}(t_1) = 0$, the atomic state at time $t_1+\tau/2$ can be determined as follows:
\begin{align*}
	C_{\estadoe}\left(t_1+\frac{\tau}{2}\right) &= \frac{e^{-i\delta\tau/4}}{\sqrt{2}}\left[- \, i \, e^{-i(\delta t_1 + \phi_1)} \right]; \\
			C_{\estadog}\left(t_1+\frac{\tau}{2}\right) &= \frac{e^{+i\delta\tau/4}}{\sqrt{2}}.
\end{align*}
Assuming that the amplitude of the ground state at time $t_2$ ($C_{\estadog}(t_2)$) is equal to the amplitude of the ground state at time $t_1 + \frac{\tau}{2}$, and similarly, the amplitude of the excited state at time $t_2$ ($C_{\estadoe}(t_2)$) is equal to the amplitude of the excited state at time $t_1 + \frac{\tau}{2}$, we can deduce the following expressions for the excited state at time $t_2 + \tau$:	
\begin{align*}
	C_{\estadoe}(t_2+\tau) &= - i\,  \frac{e^{-i\delta\tau/4} e^{-i(\delta t_2 + \phi_2)}}{\sqrt{2}} \\
	C_{\estadog}(t_2+\tau) &= - \frac{e^{+i\delta\tau/4}}{\sqrt{2}}\left[ \, e^{-i(\delta t_1 + \phi_1)}e^{i(\delta t_2 + \phi_2)} \right] 
\end{align*} 
Similarly, considering that the amplitudes of the ground state at time $t_3$ ($C_{\estadog}(t_3)$) and the excited state at the same time ($C_{\estadoe}(t_3)$) are equal to the corresponding amplitudes at time $t_2 + \tau$, namely $C_{\estadog}(t_2 + \tau)$ and $C_{\estadoe}(t_2 + \tau)$, respectively, it follows that:
\begin{align*}
	C_{\estadoe}\left(t_3+\frac{\tau}{2}\right) &= -\frac{ie^{-\delta\tau/2}}{2}\left[  1 - e^{+\delta\tau/2} e^{-i\Delta\phi} \right].
\end{align*}

Finally, the probability of measuring the excited state, given by the amplitude $C_{\estadoe}(t_3 + \frac{\tau}{2})$, is given by:
\begin{align}
	\left|C_{\estadoe}\left(t_3+\frac{\tau}{2}\right)\right|^2 &= \frac{1}{2}\left[  1 - \cos\left(\Delta\phi-\delta\tau/2 \right)\right]
    \label{eq:ProbExcitado}
\end{align}
where $\Delta\phi = \phi_1 - 2\phi_2 + \phi_3 $. Note that \eqref{eq:ProbExcitado} shows that the probability of the atom being in the excited state at the end of the $\pi/2-\pi-\pi/2$ pulse sequence depends on the phases associated with each of these pulses ($\phi_1$, $\phi_2$ and $\phi_3$).

To summarize, the primary objective of this section was to introduce, in a possibly pedagical manner, the basic physical principles governing the atom–light interaction. A two-level model was presented to elucidate the physical and mathematical concepts involved in the description of the interaction between light and atoms, which helps to clarify how laser pulses induce transitions between the internal states of atoms. However, the two-level framework does not take into account the kinetic energy of the atom nor the spatial dependence of the electric field and therefore does not fully describe the operation of a quantum gravimeter. In atomic interferometers, the motion of the atom as a whole is essential, since the laser field generates position-dependent phases and transfers momentum to the atoms. A complete understanding of the quantum gravimeter requires the inclusion of the kinetic energy term in the Hamiltonian, which is accomplished in the three-level model presented in the following section.

\section{Three-level system}\label{sec:threeLevel}
Let us consider a three-level system composed of two energy states ($\ketg$ and $\kete$) and a metastable state ($\keti$). The atom is initially in the ground state $\ketg$ with momentum $\mathbf{p}$.

Upon interacting with a pair of counter-propagating laser beams described by the electric field
\begin{equation}
	\mathbf{E} = \mathbf{E}_1 \, \cos\left(\mathbf{k}_1 \cdot \mathbf{x} - \omega_1 t + \phi_1 \right) + \mathbf{E}_2 \, \cos\left(\mathbf{k}_2 \cdot \mathbf{x} - \omega_2 t + \phi_2 \right)
    \label{eq:campoE3niveis}
\end{equation}
the atom absorbs a photon and is transiently driven to the intermediate state $\keti$ (with momentum $\mathbf{p}+\hbar\mathbf{k}_1$) and subsequently emits a photon with momentum 
$
\hbar\mathbf{k}_2$, while simultaneously reaching the excited state $\kete$ with momentum $\mathbf{p}+\hbar(\mathbf{k}_1 - \mathbf{k}_2)$. For counter-propagating lasers, we assume that the wave vectors satisfy the condition $\mathbf{k}_1 \approx - \mathbf{k}_2$. As a result, the effective wavevector of the photon interaction is given by
\begin{equation*}
    \mathbf{k}_{\text{eff}} = \mathbf{k}_1 - \mathbf{k}_2 \approx 2\mathbf{k}_1.
\end{equation*}

\begin{figure}[H]
	\centering
    \includegraphics[width=.35\textwidth]{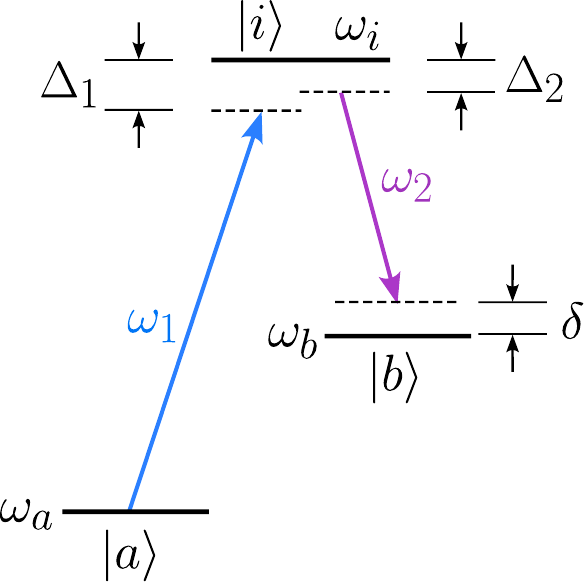}
	\caption{Schematic representation of a three-level system.}
	\label{fig:3levels}
\end{figure}

Each atomic state can be described as a tensor product between the Hilbert space of the internal degrees of freedom (energy levels) and the Hilbert space of the external degrees of freedom (momentum):
\begin{equation*}
	\mathcal H_{\text{total}}=\mathcal H_{\text{internal}}\otimes \mathcal H_{\text{external}}
\end{equation*}
and, according to the orthonormality of the internal bases, we have \citep{YOUNG1997}
\begin{align*}
 	\ketgpt &= \lvert \estadog \rangle \otimes \lvert \mathbf{p}\rangle, \\ \ketipt &= \lvert \estadoi \rangle \otimes \ketpithree,
 	\\
 	\ketephk &= \lvert \estadoe \rangle \otimes \ketpethree.
\end{align*}

The presence of three energy levels is not the only difference with respect to the previous section. In the present case, the e.m. field depends explicitly on position, as evidenced by the terms proportional to $\mathbf{k}_1\cdot \mathbf{x}$ and $\mathbf{k}_2\cdot \mathbf{x}$ appearing in Eq. \eqref{eq:campoE3niveis}. As a result, the Hamiltonian describing the system becomes
\begin{equation}
	\hat{H} = \frac{\hat{\mathbf{p}}^2}{2m} + \, \hbar \wg \projg \, + \hbar \wi \proji \,  + \hbar \we \proje - \,
	\mathbf{d}\cdot\mathbf{E}.
    \label{eq:hamiltonianthreelevels}
\end{equation}
This Hamiltonian includes two additional terms compared with the two-level case: the kinetic contribution $\hat{\mathbf{p}}^2/2m$, and the term $\hbar \wi \proji$". The former represents the kinetic energy of the particle, with momentum operator $\hat{\mathbf{p}}$ and mass $m$, whereas the latter corresponds to the internal energy associated with the metastable state $\keti$.

The quantum state at time $t$ can be written as:
\begin{align*}
	\lvert \psi(t) \rangle = \, &\Cgpt \, e^{-i \wgp t} \ketgpt + \\
    & + \Cipt \, e^{-i \wip t} \ketipt + \\
    & + \Cept \, e^{-i \wep t} \ketept.
\end{align*}
Multiplying $\lvert \dot{\psi}(t) \rangle$ by $i\hbar$ and caluclating its time derivative one gets
\begin{align*}
	i\hbar \frac{\partial}{\partial t}\lvert \psi(t) \rangle = \, &i\hbar\, \Cgptdot \, e^{-i \wgp t} \ketgpt + \\
    & + i\hbar\, \Ciptdot \, e^{-i \wip t} \ketipt + \\
    & + i\hbar\, \Ceptdot \, e^{-i \wep t} \ketept + \\
    & + \left( \hat{H}_0  + \frac{ \hat{\mathbf{p}}^2}{2m} \right) \, \lvert \psi(t) \rangle,
\end{align*}
where $\hat{H}_0$ is given by \eqref{eq:H02niveis}. 
Taking into account Eq. \eqref{eq:hamiltonianthreelevels} and using the Schr{\"o}dinger equation, one finds
\begin{align*}
	\hat{H}_{\text{int}}|\psi(t)\rangle =& \, i\hbar\, \Cgptdot \, e^{-i \wgp t} \ketgpt + \\
    & + i\hbar\, \Ciptdot \, e^{-i \wip t} \ketipt + \\
    & + i\hbar\, \Ceptdot \, e^{-i \wep t} \ketept.
\end{align*}

By multiplying the previous equation by $\langle \estadog \lvert$, $\langle \estadoe \lvert$ and $\langle \estadoi \lvert$, one obtains, respectively:
\begin{align}
	\langle \estadog \lvert \hat{H}_{\text{int}}|\psi(t)\rangle =& \, i\hbar\, \Cgptdot \, e^{-i \wgp t} \ketp \label{eq:aHPsi}\\
    \langle \estadoi \lvert \hat{H}_{\text{int}}|\psi(t)\rangle =& \, i\hbar\, \Ciptdot \, e^{-i \wip t} \ketpithree \label{eq:iHPsi}\\
    \langle \estadoe \lvert \hat{H}_{\text{int}}|\psi(t)\rangle =& \, i\hbar\, \Ceptdot \, e^{-i \wep t} \ketpethree \label{eq:bHPsi}
\end{align}

The single-photon detunings $\Delta_1$ and $\Delta_2$, corresponding to the frequencies $\omega_1$ and $\omega_2$ (Figure \ref{fig:3levels}), are defined as \citep{chu2001, Tinsley2019}:
\begin{align}
    \Delta_1 \equiv & \omega_1 - \left(\wi - \wg \right) +  \frac{\lvert \mathbf{p} \lvert^2 - \lvert \mathbf{p} + \hbar\mathbf{k}_1 \lvert^2}{2m\hbar},  \label{eq:Delta1} \\
    \Delta_2 \equiv & \omega_2 - \left(\wi - \we \right) + \frac{\lvert \mathbf{p} + \hbar\left(\mathbf{k}_1 - \mathbf{k}_2\right) \lvert^2 - \lvert \mathbf{p} + \hbar\mathbf{k}_1 \lvert^2}{2m\hbar}. \label{eq:Delta2}
\end{align}

Applying -- as in the previous section -- the RWA, which disregards the terms with rapidly oscillating exponentials in Eqs. \eqref{eq:aHPsi}, \eqref{eq:iHPsi} and \eqref{eq:bHPsi}, we obtain the following differential equations:
\begin{align}
   i \Cgptdot =& \, \frac{\Omega_{\estadog\estadoi}^{*}}{2} \Cipt \, e^{+i\Delta_1 t} e^{-i\phi_1} \label{eq:cdot_cg}\\
   \, i \Ceptdot =& \, \frac{\Omega_{\estadoe\estadoi}^{*}}{2} \Cipt \, e^{+i\Delta_2 t} e^{-i\phi_2} \label{eq:cdot_ce} \\
   i \Ciptdot =& \, \frac{\Omega_{\estadog\estadoi}}{2} \Cgpt \, e^{-i\Delta_1 t} e^{+i\phi_1} + \frac{\Omega_{\estadoe\estadoi}}{2} \Cept \, e^{-i\Delta_2 t} e^{+i\phi_2}  \label{eq:cdot_ci}
\end{align}
where $\Omega_{\estadoe\estadoi} = \frac{\langle \estadoi\lvert-\mathbf{d} \cdot \mathbf{E}_2 \lvert \estadoe \rangle}{\hbar}$ and $\Omega_{\estadog\estadoi} = \frac{\langle \estadoi\lvert-\mathbf{d} \cdot \mathbf{E}_1 \lvert \estadog \rangle}{\hbar}$.

If we assume that the amplitude terms decay much more slowly than the exponential terms, then the amplitudes can be considered time-independent and, therefore, removed from the integral \citep{Tinsley2019}:
\begin{align}
   i \int_{t_0}^{t} \Ciptdot dt' =& \frac{\Omega_{\estadog\estadoi}}{2} \Cgpt \, e^{+i\phi_1} \int_{t_0}^{t} e^{-i\Delta_1 t} + \frac{\Omega_{\estadoe\estadoi}}{2} \Cept \, e^{+i\phi_2} \int_{t_0}^{t} e^{-i\Delta_2 t} dt'. \nonumber
\end{align}
Integrating, we obtain:
\begin{align}
    i \Ceptdot =& \frac{ \Omega_{\estadog\estadoi} \Omega_{\estadoe\estadoi}^{*}}{4\Delta_1} \left( e^{-i(\Delta_1 - \Delta_2) t} - e^{i\Delta_2 t} \right) e^{i\phi} + \frac{\lvert \Omega_{\estadoe\estadoi} \lvert^2}{4\Delta_2} \left( 1 - e^{i\Delta_2 t} \right)\Cept \label{eq:ci}
\end{align}

By substituting \eqref{eq:ci} into Eqs. \eqref{eq:cdot_cg} and \eqref{eq:cdot_ce}, the latter become independent of the coefficient $\Cipt$, resulting in a system of coupled differential equations depending on two variables:
($\Cgpt$ and $\Cept$):
\begin{align}
    i \Cgptdot =& \frac{\lvert \Omega_{\estadog\estadoi} \lvert^2}{4\Delta_1} \left( 1 - e^{i\Delta_1 t} \right)\Cgpt \, + \, \frac{ \Omega_{\estadoe\estadoi} \Omega_{\estadog\estadoi}^{*}}{4\Delta_2} \left( e^{i(\Delta_1 - \Delta_2) t} - e^{i\Delta_1 t} \right) e^{-i\left(\phi_1 - \phi_2\right)} \Cept \label{eq:cdote1} \\
    i \Ceptdot =& \frac{ \Omega_{\estadog\estadoi} \Omega_{\estadoe\estadoi}^{*}}{4\Delta_1} \left( e^{-i(\Delta_1 - \Delta_2) t} - e^{i\Delta_2 t} \right) e^{i\left(\phi_1 - \phi_2\right)} + \frac{\lvert \Omega_{\estadoe\estadoi} \lvert^2}{4\Delta_2} \left( 1 - e^{i\Delta_2 t} \right)\Cept \label{eq:cdotg1}
\end{align}
Unlike the case discussed in Section \ref{section:2levels}, the detuning $\delta$ is a function of momentum and can be defined as \citep{YOUNG1997, Tinsley2019}:
\begin{equation*}
    \delta = \Delta_1 - \Delta_2 = \omega_1 - \omega_2 + \left( \omega_{\estadoe\estadog} + \frac{\mathbf{p}\cdot\mathbf{k}_{\text{eff}}}{m} + \frac{\hbar\lvert\mathbf{k}_{\text{eff}}\lvert^2}{2m} \right).
\end{equation*}
Using this definition, and applying the RWA once again, the expressions \eqref{eq:cdote1} and \eqref{eq:cdotg1} become:
\begin{align}
    i \Ceptdot =& \frac{\Omega_{\estadog\estadoi} \Omega_{\estadoe\estadoi}^{*}}{4\Delta_1} e^{-i\delta t} e^{i\left(\phi_1 - \phi_2\right)} + \frac{\lvert \Omega_{\estadoe\estadoi} \lvert^2}{4\Delta_2}\Cept \label{eq:cdote2} \\
    i \Cgptdot =& \frac{\lvert \Omega_{\estadog\estadoi} \lvert^2}{4\Delta_1} \Cgpt \, + \, \frac{ \Omega_{\estadoe\estadoi} \Omega_{\estadog\estadoi}^{*}}{4\Delta_2}  e^{i\Delta_1 t} e^{-i\left(\phi_1 - \phi_2\right)} \Cept \label{eq:cdotg2}
\end{align}

For nearly-resonant Raman transitions, the terms $\Delta_1$ and $\Delta_2$ become approximately equal, as both laser interactions involve very close internal levels and small momentum differences. Therefore, $\Delta_1 \approx \Delta_2 = \Delta$. With this approximation, 
Eqs. \eqref{eq:cdote2} and \eqref{eq:cdotg2} can be written in matrix form as
\begin{align}
	i\hbar \begin{pmatrix}
		\Ceptdot \\
		\\
		\Cgptdot
	\end{pmatrix} = \hbar\begin{pmatrix}
		\Omega_{\estadoe}^{AC} & \Omega_{\text{eff}} \, e^{-i\left(\delta\,t + \phi\right)}/2\\
		&  \\
		\Omega_{\text{eff}}^{*} \, e^{+i\left(\delta\,t + \phi\right)}/2 & \Omega_{\estadog}^{AC}
	\end{pmatrix}\begin{pmatrix}
		\Cept \\
		\\
		\Cgpt
	\end{pmatrix},
    \label{eq:sistemaCeCg}
\end{align}
with the parameters defined as:
\begin{align*}
    \Omega_{\estadoe}^{AC} &= \frac{ \Omega_{\estadog\estadoi} \Omega_{\estadoe\estadoi}^{*}}{4\Delta} \\
    \Omega_{\estadog}^{AC} &= \frac{ \Omega_{\estadoe\estadoi} \Omega_{\estadog\estadoi}^{*}}{4\Delta_2} \\
    \Omega_{\text{eff}} &= \frac{ \Omega_{\estadog\estadoi} \Omega_{\estadoe\estadoi}^{*}}{4\Delta} \\
    \phi &= \phi_2 - \phi_1
\end{align*}

Therefore, from Eq. \eqref{eq:sistemaCeCg}, the interaction Hamiltonian ($\hat{H}_{\text{int}}$) can be written as:
\begin{align}
	\hat{H}_{\text{int}} = \hbar\begin{pmatrix}
		\Omega_{\estadoe}^{AC} & \Omega_{\text{eff}} \, e^{-i\left(\delta\,t + \phi\right)}/2\\
		&  \\
		\Omega_{\text{eff}}^{*} \, e^{+i\left(\delta\,t + \phi\right)}/2 & \Omega_{\estadog}^{AC}
	\end{pmatrix}.
    \label{Hint_59}
\end{align}
Subtracting a constant energy term $(\Omega_{\estadoe}^{AC} + \Omega_{\estadog}^{AC})/2$ (of course multiplied by the $2 \times 2$ identity matrix) from the Hamiltonian \eqref{Hint_59}, one gets:
\begin{align}
	\hat{H}_{\text{int}} = \frac{\hbar}{2}\begin{pmatrix}
		(\Omega_{\estadoe}^{AC} - \Omega_{\estadog}^{AC}) & \Omega_{\text{eff}} \, e^{-i\left(\delta\,t + \phi\right)}\\
		&  \\
		\Omega_{\text{eff}}^{*} \, e^{+i\left(\delta\,t + \phi\right)} & -(\Omega_{\estadoe}^{AC} - \Omega_{\estadog}^{AC})
	\end{pmatrix}.
\end{align}
Defining $\delta^{AC} = \Omega_{\estadoe}^{AC} - \Omega_{\estadog}^{AC}$, and applying the rotating frame transformation given by \eqref{eq:rotatingH_R}, we get:
\begin{align}
	\hat{H}_R = \frac{\hbar}{2}\begin{pmatrix}
		 -\left(\delta -\delta^{AC}\right) &  & \Omega_{\text{eff}} \, e^{-i\phi}\\
		&  \\
		\Omega_{\text{eff}}^{*} \, e^{+i\phi} &  & \left(\delta -\delta^{AC}\right)
	\end{pmatrix}.
\end{align}
The eigenvalues of this matrix are given by:
\begin{align}
    \lambda \equiv  \pm\frac{\hbar\Omega_R}{2} \equiv \pm \frac{\hbar \sqrt{|\Omega_{\text{eff}}|^2 + (\delta - \delta^{AC})^2}}{2}
\end{align}
Proceeding as in Section \ref{section:2levels} and using Eq. \eqref{eq:psirot}, we obtain the following equations for the time evolution of $\Ceptt$ and $\Cgpt$:
\begin{align}
    \Ceptt(t_0+\tau) = e^{-i\left(\Omega_{\estadoe}^{AC} + \Omega_{\estadog}^{AC}\right)\tau/2} \, e^{-i\delta\tau/2}\bigg\{  &\Ceptt(t_0) \left[ \cos\left(\frac{\Omega_R\tau}{2}\right) - i\cos\theta\,\sin\left( \frac{\Omega_R\tau}{2} \right) \right] + \nonumber\\
    &+\Cgpt(t_0) e^{-i(\delta t_0 + \phi)} (-i)\sin\theta\sin\left(\frac{\Omega_R\tau}{2}\right) \bigg\};\\
	\Cgpt(t_0+\tau) = e^{-i\left(\Omega_{\estadoe}^{AC} + \Omega_{\estadog}^{AC}\right)\tau/2} \, e^{+i\delta\tau/2}\bigg\{  &\Ceptt(t_0) \left[ -i e^{i(\delta t_0 + \phi)} \sin\theta\sin\left(\frac{\Omega_R\tau}{2}\right) \right] + \nonumber \\
    &+\Cgpt(t_0) \left[ \cos\left(\frac{\Omega_R\tau}{2}\right) + i\cos\theta\,\sin\left( \frac{\Omega_R\tau}{2} \right) \right] \bigg\}.
\end{align}
Therefore, for a $\pi$-pulse of duration $\tau$, we obtain:
\begin{align}
    \Ceptt(t_0+\tau) &= -i e^{-i\left(\Omega_{\estadoe}^{AC} + \Omega_{\estadog}^{AC}\right)\tau/2} \, e^{-i\delta\tau/2}e^{-i(\delta t_0 + \phi)} \Cgpt(t_0),\\
	\Cgpt(t_0+\tau) &= -i e^{-i\left(\Omega_{\estadoe}^{AC} + \Omega_{\estadog}^{AC}\right)\tau/2} \, e^{+i\delta\tau/2} \, e^{i(\delta t_0 + \phi)}  \Ceptt(t_0),
\end{align}
while for a $\pi/2$-pulse of duration $\tau/2$ one gets:
\begin{align}
    \Ceptt(t_0+\tau/2) = e^{-i\left(\Omega_{\estadoe}^{AC} + \Omega_{\estadog}^{AC}\right)\tau/2} \, e^{-i\delta\tau/2}\bigg[ &\Ceptt(t_0) \, - \,i\Cgpt(t_0) e^{-i(\delta t_0 + \phi)}\bigg]/\sqrt{2},\\ 
	\Cgpt(t_0+\tau/2) = e^{-i\left(\Omega_{\estadoe}^{AC} + \Omega_{\estadog}^{AC}\right)\tau/2} \, e^{+i\delta\tau/2}\bigg[ &-i\Ceptt(t_0)  e^{i(\delta t_0 + \phi)} + \Cgpt(t_0) \bigg]/\sqrt{2}.
\end{align}

Assuming that at time $t_1$ we have $\Cgpt(t_1) = 1$ and $\Ceptt(t_1) = 0$, and considering that the first $\pi/2$ pulse has duration $\tau/2$, acting from $t_1$ to $t_1 + \tau/2$, we obtain the time evolution of $\Ceptt$ and $\Cgpt$ after the application of this pulse:
\begin{align}
    \Ceptt(t_1+\tau/2) &= -i e^{-i\left(\Omega_{\estadoe}^{AC} + \Omega_{\estadog}^{AC}\right)\tau/4} \, e^{-i\delta\tau/4}e^{-i(\delta t_1 + \phi_1)}/\sqrt{2},\\ 
	\Cgpt(t_1+\tau/2) &= e^{-i\left(\Omega_{\estadoe}^{AC} + \Omega_{\estadog}^{AC}\right)\tau/4} \, e^{+i\delta\tau/4} /\sqrt{2}.
\end{align}

If we assume that no additional phase shift is accumulated along the propagation of the pulses (as we will discuss in the next section), then at time $t_2$ we have $\Ceptt(t_2) = \Ceptt(t_1 + \tau/2)$ and $\Cgpt(t_2) = \Cgpt(t_1 + \tau/2)$. Therefore, the time evolution of the amplitudes after the second $\pi/2$ pulse, which takes place between $t_2$ and $t_2 + \tau$, is given by:
\begin{align}
    \Ceptt(t_2+\tau) &= -i e^{-i\left(\Omega_{\estadoe}^{AC} + \Omega_{\estadog}^{AC}\right)3\tau/4} \, e^{-i\delta\tau/4}e^{-i(\delta t_2 + \phi_2)}/\sqrt{2},\\
	\Cgpt(t_2+\tau) &= - \ e^{-i\left(\Omega_{\estadoe}^{AC} + \Omega_{\estadog}^{AC}\right)3\tau/4} \, e^{+i\delta\tau/4} \, e^{i(\delta t_2 + \phi_2)}e^{-i(\delta t_1 + \phi_1)}/\sqrt{2}. 
\end{align}
Finally, considering the last $\pi/2$ pulse, with duration $\tau/2$, applied from $t_3$ to $t_3 + \tau/2$, we obtain the final evolution of $\Ceptt$ after this pulse:
\begin{align}
    \Ceptt(t_3+\tau/2) &= -\frac{i}{2}e^{-i\left(\Omega_{\estadoe}^{AC} + \Omega_{\estadog}^{AC}\right)\tau} \, e^{-i\delta\tau/2}\bigg[1 - e^{+i\delta\tau/2} \, e^{\Delta\phi}\bigg].
\end{align}
It follows that the probability of finding the system in the excited state ($\Ceptt$) after the third pulse is given by:
\begin{align}
	\left|\Ceptt(t_3+\tau/2)\right|^2 &= \frac{1}{2}\left[  1 - \cos\left(\Delta\phi-\delta\tau/2 \right)\right].
    \label{eq:ProbExcitadoThree}
\end{align}

It is noteworthy that Eq. \eqref{eq:ProbExcitadoThree} is identical to \eqref{eq:ProbExcitado}, obtained in the two-level case (Section \ref{section:2levels}). This indicates that the population probabilities in the three-level system in the considered approximations the same as those predicted by the two-level approximation.

\section{Gravity Measurement}

In the previous section, it was shown that the probability of detecting the atom in the excited state after the sequence of $\pi/2-\pi-\pi/2$ pulses is given by Eq. \eqref{eq:ProbExcitadoThree}. Up to this point, the effect of gravitational acceleration had not been taken into account. Therefore, the interferometric phase $\Delta \phi$ is determined by the phases ($\phi_1$, $\phi_2$ and $\phi_3$) of the $\pi/2-\pi-\pi/2$ pulse sequence. The final contribution -- in absence of gravity (i.e., $g=0$) -- arises from the following sum of the phases generated by the three laser pulses:
\begin{equation}
    \Delta\phi_{\text{laser}} = \phi_1 - 2\phi_2 + \phi_3.
    \label{DeltaPhi_75}
\end{equation}

However, in the presence of a gravitational potential, the particle undergoes a continuous variation in its transition frequency, which introduces an additional phase shift at the output of the interferometer. In other words, the total measured phase ($\Delta\phi_{\text{tot}}$) does not depend only on the pulse phases $\phi_1$, $\phi_2$, and $\phi_3$ and there is a phase shift $\Delta\phi_{\text{grav}}$ induced by the action of gravity on the trajectory of the particle. Thus, we write for pedagogical reasons the total phase as
\begin{equation}
    \Delta\phi_{\text{tot}} = \Delta\phi_{\text{laser}} + \Delta\phi_{\text{grav}},
    \label{DeltaPhi_tot}
\end{equation}
where $\Delta\phi_{\text{laser}}$ is given by \eqref{DeltaPhi_75}.

\begin{figure}[H]
	\centering
    \includegraphics[width=.65\textwidth]{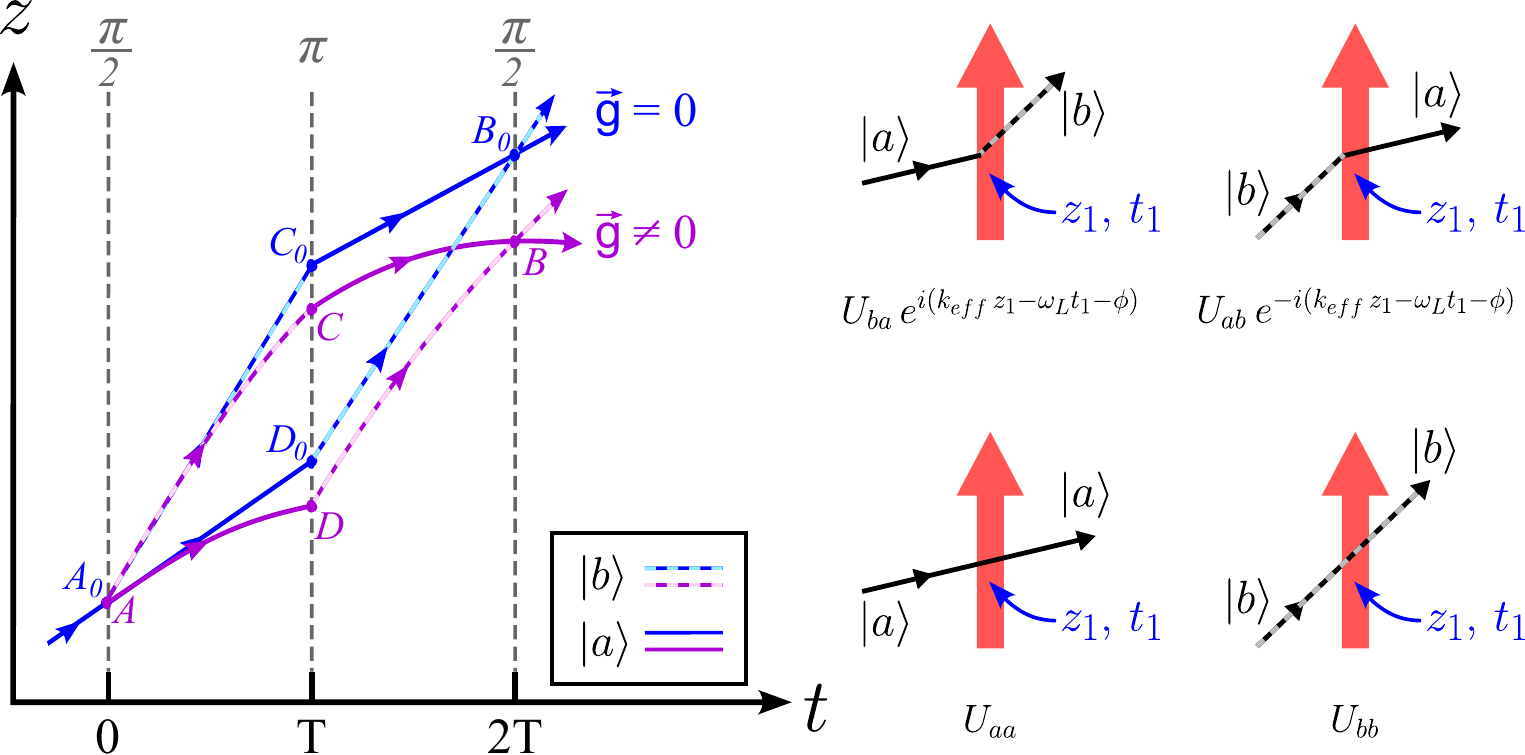}
	\caption{Left: an illustration is shown of the trajectory described by the atoms in the interferometer, subjected to the $ \pi/2 - \pi - \pi/2$ pulse sequence, considering the cases of absence (blue lines) and presence (purple lines) of a gravitational field. Right: a representation is provided of the four types of interaction between an atom and the laser beam (red arrows) with frequency $\omega_L$ and wavevector $k_{\text{eff}}$ (this interaction occurring at a time $t_1$ in a position $z_1$): (top left) the atom absorbs a photon, gaining momentum $\hbar k_{\text{eff}}$ and transitioning from state $\ketg$ to state $\kete$; (top right) the atom emits a photon, losing momentum $\hbar k_{\text{eff}}$ and decaying from state $\kete$ to state $\ketg$; (bottom left) the atom remains in state $\ketg$; (bottom right) the atom remains in state $\kete$.}
	\label{fig:transicoes}
\end{figure}

To understand how to determine 
$\Delta\phi_{\text{grav}}$ we can proceed in the following way. 
In the absence of gravity, the trajectories described by the atoms are straight lines, since they maintain constant velocity in each arm of the interferometer. The only velocity changes arise from the absorption or emission of photons during the Raman pulses. Nevertheless, in the presence of a gravitational field, the atomic velocities vary continuously along their trajectories due to the gravitational acceleration experienced by the atoms. As a consequence, the paths become curved in the ``depth $\times$ time" diagram (Figure \ref{fig:transicoes}). This curvature introduces an additional phase shift, denoted by $\Delta\phi_{\text{path}}$ \citep{Storey1994, bhardwaj2024}, which can be determined by
\begin{equation}
    \Delta\phi_{\text{path}} = \frac{S_{cl}}{\hbar},
    \label{DeltaPhiPath}
\end{equation}
where $S_{cl}$ represents the action evaluated along the classical trajectories \citep{chu2001}. A derivation of \eqref{DeltaPhiPath} is in Appendix \ref{secA2} and in the next discussion we closely follow \citet{Storey1994}.

The classical action from a point $z_1$ (at a time $t_1$) to a point $z_2$ (at a time $t_2$) is defined as the time integral of the Lagrangian $L(z,\dot z)$ along the atomic trajectory:
\begin{align}
	S_{cl} \left(z_2 \, t_2, z_1 \, t_1\right) =& \int_{t_1}^{t_2}  L(z,\dot{z})\, dt. \nonumber
\end{align}
The Lagrangian $L(z,\dot{z})$ is written as
\begin{equation*}
    L(z(t),\dot{z}(t)) = \frac{1}{2}m \dot{z}(t)^2 - mgz(t),
\end{equation*}
where $z(t)$ and $\dot{z}(t)$ are the equations of motion of the particle, which initially (at time $t_1$) has velocity $v_1$ at position $z_1$:
\begin{align*}
    z(t) &= z_1 + v_1 (t - t_1) - \frac{1}{2}g(t-t_1)^2, \\
    \dot{z}(t) &= v_1 - g(t-t_1).
\end{align*}
Based on this, the solution for the classical action $S_{cl} \left(z_2 \, t_2, z_1 \, t_1\right)$, is given by \citep{Storey1994}
\begin{align}
	S_{cl} \left(z_2 \, t_2, z_1 \, t_1\right) =&   \, \frac{m}{2}\frac{(z_2 - z_1)^2}{(t_2 - t_1)} \, - \, \frac{m \, g}{2} \left(t_2 -t_1\right) \left(z_2 + z_1\right) \, - \, \frac{m g^2}{24} \left(t_2 - t_1\right)^3. \label{eqAction}
\end{align}
Therefore, the total contribution to the phase difference between the two arms due to propagation, according to the geometry described in the left part of Figure \ref{fig:transicoes} is 
\citep{Storey1994}:
\begin{align*}
	\Delta \phi_{\text{path}} &= \left\{ \left[S_{cl}\left(AC\right) - S_{cl}\left(AD\right) \right]+ \left[S_{cl}\left(CB\right) - S_{cl}\left(DB\right)\right]\right\}/\hbar \\
	&= \frac{m}{T\hbar}\left(z_C - z_D\right) \left[z_C + z_D - z_A - z_B - g \,  T^2 \right] 
\end{align*}
Notice that we are denoting with 
$z_A, z_B, z_C, z_D$ the positions in which the atom is at the different pulses according their location in 
Figure \ref{fig:transicoes}, while the corresponding quantities in absence of gravity with a further index $_{_0}$, also indicated in  Figure \ref{fig:transicoes}. 
By geometric considerations
\begin{equation*}
	z_{D_0}-z_D = (1/2) gT^2 = z_{C_0}-z_C, 
\end{equation*}
\begin{equation*}
	z_{B_0}-z_B = 2gT^2. 
\end{equation*}
Moreover, since $A_0 B_0 C_0 D_0$ is a parallelogram, this implies that the phase difference along the two arms in absence of gravity has to be vanishing. It follows 
\begin{equation*}
	z_C + z_D - z_A - z_B - g \,  T^2 = z_{C_0} + z_{D_0} - z_{A_0} - z_{B_0} = 0,
\end{equation*}
from which $\Delta\phi_{\text{path}}=0$ and
\begin{equation}
	g \,  T^2 = z_C + z_D - z_A - z_B.
    \label{eq:gT2}
\end{equation}
The fact that $\Delta\phi_{\text{path}}=0$ does not imply that $\Delta\phi_{\text{grav}}$, 
as defined in \eqref{DeltaPhi_tot} with $\Delta\phi_{\text{laser}}$ given by Eq. \eqref{DeltaPhi_75}, is vanishing as well, as we are going to discuss now.

Let analyze the state transitions induced by the laser pulses. The interactions between the atom and the Raman pulses can occur in four distinct ways (right part of Figure \ref{fig:transicoes}):
 \begin{itemize}
    \item The atom absorbs a photon, acquiring a momentum $\hbar k$ and undergoing a transition from the state $\lvert \estadog\rangle$ to the state $\lvert \estadoe\rangle$;
    \item The atom emits a photon, losing a momentum $\hbar k$ and decaying from the state $\lvert \estadoe\rangle$ to the state $\lvert \estadog\rangle$;
    \item The atom remains in the state $\lvert \estadog\rangle$ during the laser interaction;
    \item The atom remains in the state $\lvert \estadoe\rangle$ during the laser interaction.
\end{itemize}

These four processes determine the accumulated phase associated with each interaction with the laser pulses. The accumulated phase resulting from the pulses can be calculated from the transition operators $U_{ij}$ 
associated with each Raman pulse according Table \ref{tab:amp_fases}.
\begin{table}[h]
\caption{Amplitude and phase associated with each  transition.}\label{tab:amp_fases}%
\begin{tabular}{@{}llll@{}}
\toprule
\textbf{Time}  & \textbf{Pulse} & \textbf{Upper arm}  & \textbf{Lower arm} \\
\midrule
$t=0$    & $\pi/2$ & $U_{a a}$  & $ U_{b a} e^{+i\left(k_{\text{eff}} z_A - \phi_1\right)}$  \\
$t=T$    & $\pi$   & $ U_{b a} e^{+i\left(k_{\text{eff}} z_C - \phi_{2}\right)}$  & $ U_{a b} e^{-i\left(k_{\text{eff}} z_D - \phi_{2}\right)}$   \\
$t=2T$   & $\pi/2$ & $ U_{b b} $ & $U_{b a} e^{+i\left(k_{\text{eff}} z_B - \phi_{3}\right)}$ \\
\botrule
\end{tabular}
\end{table}

Given the information contained in Table \ref{tab:amp_fases}, the final amplitudes and phase for the upper arm are:
\begin{align*}
	& U_{a a} U_{b b}\, e^{+i\left[k_{\text{eff}}z_C - \phi_{2}\right]},
\end{align*}
while for the lower arm we obtain
\begin{align*}
	&U_{b b}\, U_{a a} \, e^{+i\left(k_{\text{eff}} \left(z_A - z_D + z_B\right) - \phi_1 + \phi_{2} - \phi_{3} \right)}.
\end{align*}
The difference between the accumulated phases in the two arms results therefore in:
\begin{align*}
	\Delta \phi_{tot} &= k_{\text{eff}} \left(z_C - z_B\right) + \omega_L T - \phi_{2} + \phi_{3} - \left[k_{\text{eff}} \left(z_A - z_D\right) + \omega_L T - \phi_1 + \phi_{2}\right] = \\
	 &= k_{\text{eff}} \left(z_C - z_B - z_A + z_D\right) + \phi_1 - 2 \phi_{2} + \phi_{3},
\end{align*}
where the presence of $k_{\text{eff}}$ 
follow from the phases in the operators $U$'s written in Table \ref{tab:amp_fases}.

Using Eq. \eqref{eq:gT2}, we arrived at the main result we were looking for: The total accumulated phase can be expressed as
\begin{align}
	\Delta \phi_{tot} &= k_{\text{eff}} g T^2 + \phi_1 - 2 \phi_{2} + \phi_{3}.
    \label{eq:phitot}
\end{align}
In other words, 
$\Delta\phi_{\text{grav}}$, 
as defined in \eqref{DeltaPhi_tot} with $\Delta\phi_{\text{laser}}$ given by Eq. \eqref{DeltaPhi_75}, is given by 
$\Delta\phi_{\text{grav}}=k_{\text{eff}} g T^2$.

The effect of gravitational acceleration acts continuously along the entire atomic trajectory, such that the term $k_{\mathrm{eff}} g T^{2}$ represents the total phase accumulated at $t = 2T$. However, gravitational acceleration also shifts the atomic resonance frequency; i.e., an atom that is initially in resonance with the first $\pi/2$ pulse will no longer remain resonant with the subsequent pulses if the laser frequencies are kept fixed.

To compensate for this effect, a linear chirp may be applied to the Raman frequency difference 
\citep{YOUNG1997}. Consequently, 
the time-dependent phase, which we denote by $\Phi(t)$, 
evaluated at each pulse is given by
\begin{align}
    \Phi_1(t_1) &= \omega_1 t_1 - k_{\mathrm{eff}} g\, t_1^{2} + \phi_{1}, \\
    \Phi_2(t_2) &= (\omega_1 + \omega_m)t_2 - k_{\mathrm{eff}} g\, t_2^{2} + \phi_{2}, \\
    \Phi_3(t_3) &= (\omega_1 + 2\omega_m)t_3 - k_{\mathrm{eff}} g\, t_3^{2} + \phi_{3}.
\end{align}
Substituting $t_1 = 0$, $t_2 = T$, and $t_3 = 2T$, one obtains
\begin{align}
    \Delta\Phi
    &= \Phi_1(0) - 2\Phi_2(T) + \Phi_3(2T) \nonumber \\
    &= 2\, \omega_m T - k_{\mathrm{eff}}\, g\, T^{2} + \Delta\phi_{\text{laser}}, \label{eq:dphif}
\end{align}
where $\omega_m$ is a frequency to be determined.

It is possible to maintain resonance by applying a single continuous chirp to the Raman frequency difference:
one obtains
\begin{equation}
\Delta \Phi = (\beta - k_{\mathrm{eff}} g) T^{2},
\end{equation}
where $\beta=\omega_m/T$. 
When the rate of variation of the Raman frequency (chirp) satisfies the condition
\[
\beta = k_{\mathrm{eff}}\, g,
\]
the total interferometric phase becomes zero, that is, $\Delta\Phi = 0$. In this situation, the Doppler shift associated with the gravitational acceleration of the atoms is completely compensated by the chirp applied to the Raman beams. Thus, the phase cancellation condition can be written as
\begin{equation}
	\beta - k_{\mathrm{eff}}\, g = 0,
\end{equation}
from which one directly obtains \citep{Menoret_2018}
\begin{equation}
	g = \frac{\beta}{k_{\mathrm{eff}}}.
	\label{eq:gbeta}
\end{equation}

The experimental parameters $\tau$, $T$, and $\Delta\phi_{0}$ are known, 
since they correspond to controlled quantities. 
With these parameters fixed, the total phase accumulated over the three-pulse Raman interferometric sequence can be inferred from the transition probability to the excited state, as described by Eqs. \eqref{eq:ProbExcitado} and \eqref{eq:ProbExcitadoThree}. Experimentally, this probability is determined by the ratio between the number of atoms detected in the excited state and the total number of atoms after the application of the three laser pulses. Therefore, by varying the frequency chirp $\beta$, the interferometric phase is modified and, consequently, the final population in the excited state is changed. In this manner, the phase associated with the interference can be continuously tuned through the control of $\beta$. The value of the local gravitational acceleration $g$ is then determined by identifying the value of $\beta$ that cancels the interferometric phase and by applying the relation given in Eq. \eqref{eq:gbeta}.

\section{Analysis of the Quantum Gravimeter Stability}

The precision and stability of quantum gravimeters rely on the ability to characterize the various noise sources that affect the phase measurement, as the signals obtained from these instruments often exhibit non-Gaussian and non-stationary noise. As seen in the previous sections, the output of an atomic interferometer is a phase accumulated by the atoms (sensitive to local gravity), but also to various sources of noise, among which the 
laser noise and phase fluctuations caused by mechanical vibrations of the experimental platform \citep{Le_Gou_t_2008}.

In this context, the Allan deviation \citep{Allan1966,riley} arises as a tool to quantify the stability of signals over different time scales. Additionally, a complementary description can be obtained in the frequency domain, where the phase variance is calculated from the power spectral density (PSD).
\begin{figure}[H]
	\centering
    \includegraphics[width=0.5\textwidth]{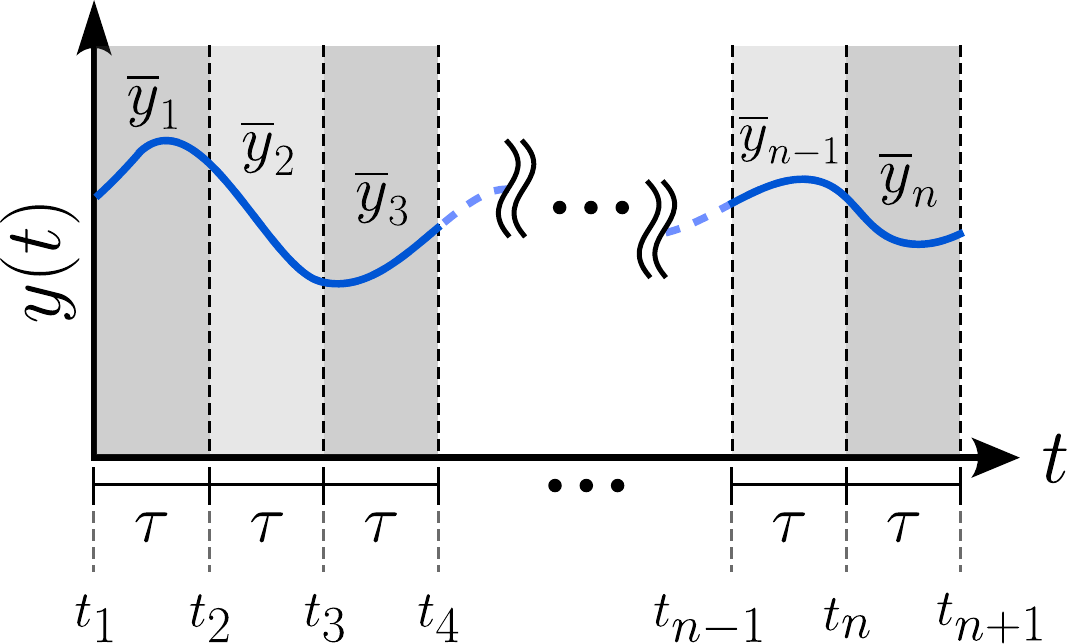}
	\caption{Illustration of the sampling of the data $y(t)$ into $n$ intervals of duration $\tau$, where, for each interval, the mean of the values contained within it is calculated. That is, for the $i$-th interval, the mean $\bar{y}_i$ of all the $y$ samples within that interval is computed.}
	\label{fig:allan_ilustration}
\end{figure}

The computation of the Allan deviation for a time series $y(t)$ is based on the statistical comparison of successive averages of the signal over windows of duration $\tau$. For a given value of $\tau$, the total observation interval is partitioned into $n$ adjacent blocks, each covering the interval $[t_i,\, t_{i+1}]$\citep{Rubiola_2008}, as shown in Figure \ref{fig:allan_ilustration}. The average of the signal in each block is then defined as
\begin{equation}
    \bar{y}_{i} (\tau) = \frac{1}{\tau} \int\limits_{t_i}^{t_{i+1}} y(t)\, dt
\end{equation}
where $i = 1,2,3,\ldots,n$. The corresponding Allan variance is given by
\begin{equation}
    \sigma_{y}^2 (\tau) = \frac{1}{2(n-1)} \sum\limits_{i=1}^{n} \big[ \bar{y}_{i+1}(\tau) - \bar{y}_{i}(\tau) \big]^2,
\end{equation}
and the Allan deviation is simply $\sigma_y = \sqrt{\sigma_y^2}$.

In practice, $y(t)$ is not continuous but sampled, such that each interval $\tau$ contains $m$ samples. Therefore, the average over the $i$-th window is given by
\begin{equation*}
    \bar{y}_i = \frac{1}{m} \sum\limits_{k=1}^{m} y_k,
\end{equation*}
where the index $k$ runs over all samples contained within the $i$-th window.

The variance $\sigma_y^2$ can be described in the frequency domain as \citep{Peters_2001, Rubiola_2008}:
\begin{equation}
\sigma_y^2 = \int_0^{\infty} H(\omega) S(\omega) \, d\omega,
\end{equation}
where $H(\omega)$ is the frequency-dependent response function of the interferometer, and $S(\omega)$ is the power spectral density of the noise sources affecting the phase measurement. By using the power spectral density $S(\omega)$, it is possible to understand how noise at different frequencies contributes to the overall phase instability, providing deeper insights into the behavior of the gravimeter.

\subsection{Sensitivity to Phase Variations} 
The final phase $\Phi$ measured in an interferometer is not acquired instantaneously, but accumulated throughout the operation of the instrument. Thus, any temporal fluctuation $\delta\phi(t)$ in the phase affects the final measured value. However, the interferometer does not respond uniformly to disturbances applied at different moments. The sensitivity to these fluctuations depends on the time at which they occur, and this temporal dependence is described by the sensitivity function $g_s(t)$.

Consider an infinitesimal perturbation $\delta\phi(t)$ applied to the phase of the Raman lasers at time $t$. This perturbation generates a corresponding variation $\delta P$ in the transition probability $P$ (given by Eqs. \eqref{eq:ProbExcitado} and \eqref{eq:ProbExcitadoThree}). The sensitivity function is defined as the limit of the ratio between these two quantities \citep{Le_Gou_t_2008, Patrick2008, Tinsley2019}:
\begin{equation} 
    g_s = 2 \lim_{\delta\phi \to 0} \frac{\delta P
    }{\delta\phi}. 
\end{equation} 

For maximum sensitivity, the interferometer is assumed to operate at the point of maximum fringe slope, i.e., under the condition in which the total phase is $\Phi = \pi/2$ \citep{Patrick2008, Tinsley2019}. In this regime, the transition probability is $P = 1/2$, and small phase perturbations generate linear and symmetric variations around the operating point. The transition probability may be written as
\begin{equation} 
    P = \frac{1}{2}\left[1 - \cos\left(\frac{\pi}{2} \pm \delta\Phi(t)\right)\right].
\end{equation} 
    
Using $\cos\left( \frac{\pi}{2} \pm  \delta\Phi \right) = \mp \sin\left( \delta\Phi \right)$ and $\sin\delta\Phi \approx \delta\Phi$, we obtain $P \approx \frac{1 \mp \delta\Phi(t)}{2}$, from  
which $\delta P = \frac{\delta\Phi(t)}{2}$ and, $g_s = \lim_{\delta\phi \to 0} \frac{\delta\Phi(t)}{\delta\phi}$. Following \cite{Patrick2008}, to which we refer for more details, one finds that $g_s$ is given by
\begin{equation}
    g_s (t) =\begin{cases} 
	    \sin(\Omega_R t), & 0 < t < T/2 \\ 
	    1, & T/2 < t < T + T/2 \\ 
	    \sin(\Omega_R (t - T)), & T + T/2 < t < T + T 
    \end{cases}
\end{equation}.

\begin{figure}[H]
	\centering
    \includegraphics[width=1.\textwidth]{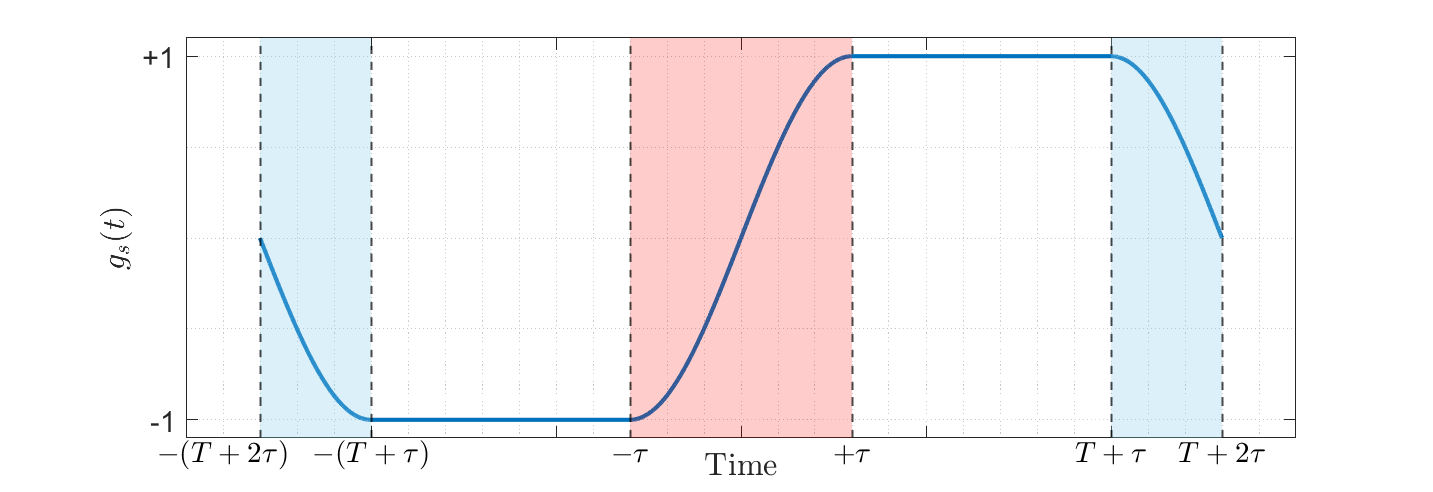}
	\caption{Sensitivity function:  The regions shaded in blue represent the intervals during the operation of the $\pi/2$ pulses, while the red region corresponds to the interval during the operation of the $\pi$ pulse.}
	\label{fig:sensibilidade}
\end{figure}

A noise component $\delta\phi$ applied between the first and the second pulses induces a variation $\delta\Phi = -\delta\phi$ in the accumulated phase $\Phi$, resulting in $\delta P \simeq -\delta\phi/2$. Consequently, the sensitivity function assumes the value $-1$ within this temporal interval, while between the second and the third pulses it assumes the value $+1$ \citep{Patrick2008}, as shown in Figure \ref{fig:sensibilidade}.

The sensitivity function establishes a direct relationship between the temporal phase fluctuations and the total phase accumulated in the interferometer. Thus, any noise source affecting the phase, whether originating from the laser, mechanical vibrations of the experimental platform, or other sources, can be incorporated into the sensitivity function to estimate its impact on the gravity measurement.

In the frequency domain, the sensitivity function also allows calculating the contribution of noise to the instability of the accumulated phase using the noise power spectral density $S(\omega)$, with the Allan variance of the accumulated phase being expressed as \citep{Le_Gou_t_2008}:
\begin{equation}
	\sigma_\Phi^2 = \int_0^{\infty} \, \Big[\omega|G(\omega)|\Big]^2 \, S_{\phi}(\omega) \, d\omega,
\end{equation}
where $G(\omega)$ is the Fourier transform of the sensitivity function $g_s(t)$. This formulation provides a powerful tool to quantify the influence of different noise components -- white noise, phase noise, mechanical vibrations, among others.

\subsection{Noise Contributions to the Quantum Gravimeter Sensitivity}

The sensitivity of a quantum gravimeter is affected by any noise source capable of generating perturbations in the measured interferometric phase. Temporal phase fluctuations couple to the total accumulated phase through the sensitivity function, resulting in a phase variance that degrades the precision of the gravity measurement. This formalism provides a unified framework to describe the influence of different noise sources, both internal and external to the interferometric system.

A particularly relevant example is the noise arising from mechanical vibrations of the platform on which the gravimeter is installed. When the platform vibrates, the freely falling atoms do not follow this motion. However, when adopting the reference frame defined by the interferometer structure -- in particular, the retroreflection mirror of the Raman beams -- such vibrations are equivalent to an acceleration imposed on the atoms. Consequently, this relative motion between the atoms and the interferometer reference frame manifests itself as an effective acceleration fluctuation, which couples directly to the measured interferometric phase \citep{miffre2006}.

As described in \cite{Patrick2008}, this effect can be modeled by introducing a spurious acceleration $\delta a(t)$, which couples to the interferometric phase through the effective wave vector $k_{\mathrm{eff}}$. The corresponding variation of the accumulated phase can then be written as
\begin{equation}
    \delta\Phi = k_{\mathrm{eff}} \int g_a(t)\, \delta a(t)\, dt,
\end{equation}
where $g_a(t)$ is the interferometer sensitivity function to accelerations. This function is directly related to the phase sensitivity function $g_s(t)$ through the relation
\begin{equation}
    g_a(t) = \frac{1}{k_{\mathrm{eff}}}\,\frac{d^2 g_s(t)}{dt^2}.
\end{equation}

In the frequency domain, this relation allows the Allan variance associated with vibration noise to be expressed in terms of the power spectral density of the acceleration noise $S_a(\omega)$. For an integration time $\tau$, the resulting Allan variance can be written as \citep{Le_Gou_t_2008, Patrick2008}
\begin{equation}
    \sigma_a^2(\tau) =
    \frac{k_{\mathrm{eff}}^2}{\tau_m}
    \int_0^{\infty}
    \left| \frac{G(\omega)}{\omega^2} \right|^2
    S_a(\omega)\, d\omega,
    \label{eq:allan_vibration_noise}
\end{equation}
where the factor $1/\omega^2$ strongly suppresses the contribution of high-frequency vibrations. As a consequence, low-frequency components of $S_a(\omega)$ dominate the measurement variance, while high-frequency vibrations have a significantly reduced impact on the estimation of the gravitational acceleration.

\section{Conclusion}\label{sec13}
In this work, we present a pedagogical description of the operating principle of ultracold quantum gravimeters, with the objective of providing content specifically directed to geoscientists. We start with a brief presentation of the fundamental concepts of quantum mechanics, such as the superposition principle, the bra--ket notation, and the time evolution described by the Schr\"{o}dinger equation. These constitute the basis for the modeling of two- and three-level atomic systems interacting with laser fields.

In addition, we discuss how laser pulses can play roles analogous to splitters and mirrors for electromagnetic waves, culminating in the realization of a Mach--Zehnder--type atomic interferometer. In this context, we elucidate how the phase accumulated along the different arms of the interferometer is directly related to the gravitational acceleration and how this information can be experimentally retrieved from the final populations of the atomic states after the interferometric pulse sequence.

A central point of the analysis lies in the fact that 
for the three-level Raman scheme employed in quantum gravimeters  
the final expression for the transition probability exhibits the same functional form as that obtained in an effective two-level model. This equivalence highlights the robustness of the underlying interferometric principle and justifies the recurrent use of simplified two-level descriptions in theoretical studies and practical applications.

We further discuss how gravity contributes to the interferometric phase through the classical action associated with the atomic trajectories and how the introduction of a frequency chirp in the Raman lasers allows compensation of the Doppler shift induced by gravitational acceleration. This procedure leads to the phase dependence proportional to $k_{\mathrm{eff}} g T^2$  which constitutes the basis of absolute gravity measurements performed with atomic interferometers.

Additionally, we analyze the stability of the quantum gravimeter using the Allan variance, emphasizing its role as a statistical tool for characterizing the instrument performance over different time scales. Both time-domain and frequency-domain formulations are presented. We show how the interferometer sensitivity function acts as the link between temporal phase fluctuations and the variance of the accumulated phase, enabling a quantitative assessment of the contribution of different noise sources, with particular emphasis on mechanical vibration noise.

By presenting the theoretical foundations in a detailed manner, we intend to make this work an useful reference for geoscientists interested in understanding, employing, or improving quantum gravimeters. As quantum sensors become increasingly commercially accessible, their impact on field measurements, long-term monitoring, and gravimetric applications is expected to expand. In this context, knowledge of the underlying quantum principles 
becomes an increasingly indispensable competence.

Accordingly, we view this contribution as an invitation to the geoscience community for deeper engagement with measurement techniques based on quantum physics, fostering interdisciplinary research at the interface between fundamental physics and the Earth sciences.

\backmatter

\bmhead{Acknowledgements}
We acknowledge Fabio Benatti and Ugo Marzolino for
helpful and useful discussions. I.R.d.S.J. thank support from the Project “National Quantum Science and Technology In-
stitute – NQSTI” Spoke 3: “ Atomic, molecular platform for quantum technologies”.










\begin{appendices}

\section{Eigenvectors of $\hat{H}_{R}$}\label{secA1}

We determine here the eigenvectors of
\begin{align}
	\hat{H}_{R} = \frac{\hbar}{2}\begin{pmatrix}
		-\delta & \Omega_{\estadoe\estadog} \, e^{-i\phi}\\
		& \\
		\Omega_{\estadoe\estadog}^{*} \, e^{+i\phi} & +\delta,
	\end{pmatrix}
\end{align}
whose eigenvalues are given by $\lambda_{\pm} = \pm \frac{\hbar \Omega_R}{2}$. 

The eigenvector $\lvert \lambda_{+} \rangle = \begin{pmatrix} a \\ b \end{pmatrix}$, associated with the eigenvalue $\lambda_{+} = + \frac{\hbar \Omega_R}{2}$, satisfies the eigenvalue equation:
\begin{equation}
    \left(\hat{H}_{R} - \lambda_{+} \boldsymbol{I} \right) \lvert \lambda_{+} \rangle = \boldsymbol{0}.
\end{equation}
Thus,
\begin{align}
	\begin{pmatrix}
		-\left(\delta + \Omega_R \right)  & \Omega_{\estadoe\estadog} \, e^{-i\phi}\\
		& \\
		\Omega_{\estadoe\estadog}^{*} \, e^{+i\phi} & \left(\delta - \Omega_R \right)
	\end{pmatrix}\begin{pmatrix} a \\ \\ b \end{pmatrix} = \begin{pmatrix} 0 \\ \\ 0 \end{pmatrix}.
\end{align}
Consequently, 
\begin{equation}
    b = \frac{\Omega_R+\delta}{\Omega_{\estadoe\estadog}} \, e^{+i\phi} \, a
    \label{eq:b_append}
\end{equation}
From Eq. \eqref{eq:b_append}, we obtain
\begin{equation}
b = \sqrt{\frac{\Omega_R+\delta}{\Omega_{R} - \delta}} \, e^{+i\phi} \, a.
\end{equation}
Considering Eqs. \eqref{eq:sincos}, and using the trigonometric half-angle identities, we have:
\begin{align}
    \sin \frac{\theta}{2} \, = \, \sqrt{\frac{\Omega_R + \delta}{2 \, \Omega_{R}}},  \hspace{2cm} \cos \frac{\theta}{2} \, = \, \sqrt{\frac{\Omega_{R} - \delta}{2 \, \Omega_R}}
\end{align}
Therefore,
\begin{align}
    b = \frac{\sin \frac{\theta}{2}}{\cos \frac{\theta}{2}} \, e^{+i\phi} \, a
\end{align}
and the eigenvector $\lvert \lambda_{+} \rangle $ can be written as:
\begin{align}
    \lvert \lambda_{+} \rangle = \begin{pmatrix} a \\ \\ b \end{pmatrix} = \frac{a e^{+i\frac{\phi}{2}}}{\cos \frac{\theta}{2}}\begin{pmatrix} \cos \frac{\theta}{2} \, e^{-i\frac{\phi}{2}} \\ \\ \sin \frac{\theta}{2} \, e^{+i\frac{\phi}{2}} \end{pmatrix}
\end{align}
(with $a$ fixed by the normalization).

To determine the eigenvector  $\lvert \lambda_{-} \rangle$, we use the same procedure with the eigenvalue equation given by $\left(\hat{H}_{R} - \lambda_{-} \boldsymbol{I} \right) \lvert \lambda_{-} \rangle = \boldsymbol{0}$, 
from which 
\begin{equation}
    -b = \frac{\Omega_R - \delta}{\Omega_{\estadoe\estadog}} \, e^{+i\phi} \, a
    \label{eq:b_append}
\end{equation}
and 
\begin{align}
    -b &= \sqrt{\frac{\Omega_R - \delta}{\Omega_{R} + \delta}} \, e^{+i\phi} \, a 
    \label{eq:b_append}.
\end{align}
and $b = -\frac{\cos \frac{\theta}{2}}{\sin \frac{\theta}{2}} \, e^{+i\phi} \, a$.
Therefore, with $a$ fixed by the normalization, the eigenvector $\lvert \lambda_{-} \rangle $ can be expressed as:
\begin{align}
    \lvert \lambda_{-} \rangle = \begin{pmatrix} a \\ \\ b \end{pmatrix} = -\frac{a e^{+i\frac{\phi}{2}}}{\sin \frac{\theta}{2}}\begin{pmatrix} -\sin \frac{\theta}{2} \, e^{-i\frac{\phi}{2}} \\ \\ \cos \frac{\theta}{2} \, e^{+i\frac{\phi}{2}} \end{pmatrix}
\end{align}

\section{Phase Arising from the Classical Action ($S_{\mathrm{cl}}$)} \label{secA2}

When one considers the propagation of a particle from a position $x_a$ at time $t_a$ to a position $x_b$ at time $t_b$, the wave function associated with this motion can be described by the quantum propagator. This propagator, $K(x_b,t_b;x_a,t_a)$, contains all the information about the system and describes how the wave function evolves between the points $x_a$ and $x_b$. This evolution is governed by the evolution operator $U(t_b,t_a)$, acting on the wave function at the initial time $t_a$, it yields the wave function at the later time $t_b$:
\begin{equation*}
    \lvert\psi(x_b,t_b)\rangle=U(t_b,t_a)\,\lvert\psi(x_a,t_a)\rangle.
\end{equation*}
To project this equation onto the position basis, one applies the projection of the state $\lvert \psi(x_b,t_b) \rangle$ onto the position eigenstates $\lvert x_b \rangle$:
\begin{equation*}
    \langle x_b \lvert\psi(x_b,t_b)\rangle=\langle x_b \lvert U(t_b,t_a)\,\lvert\psi(x_a,t_a)\rangle.
\end{equation*}
For the completeness relation 
$\mathbf{I} = \int dx_a \lvert x_a \rangle \langle x_a \rvert$, the previous expression can be rewritten as
\begin{equation*}
    \langle x_b \lvert\psi(x_b,t_b)\rangle=\langle x_b \lvert U(t_b,t_a)\,\lvert x_a\rangle\langle x_a \lvert\psi(x_a,t_a)\rangle.
\end{equation*}
Defining the propagator as
\begin{equation*}
    K(x_b,t_b;x_a,t_a)=\langle x_b|U(t_b,t_a)|x_a\rangle,
\end{equation*}
the equation that expresses the wave function $\psi(x_b,t_b)$ in terms of the propagator is then given by
\begin{equation*}
    \psi(x_b,t_b)=\int dx_a\,K(x_b,t_b;x_a,t_a)\,\psi(x_a,t_a).
\end{equation*}

Within the framework of Feynman path integrals, the propagator can be interpreted as a sum over all possible paths $\Gamma$ connecting the points $(x_a,t_a)$ and $(x_b,t_b)$ \citep{Shankar,feynman2010quantum}:
\begin{equation*}
K(x_b,t_b;x_a,t_a)=\mathcal{N}\sum_{\Gamma} e^{\frac{i}{\hbar}S(\Gamma)},
\end{equation*}
where $S(\Gamma)$ denotes the classical action associated with the path $\Gamma$,
\begin{equation*}
S=\int_{t_a}^{t_b} L(x,\dot{x})\,dt,
\end{equation*}
is the action with $L$ the Lagrangian of the system, and $\mathcal{N}$ is a normalization factor.

In the 
semiclassical regime, characterized by $S\gg\hbar$, the path integral is dominated by classical trajectories. As a consequence, the evolution of the wave function can be approximated by
\begin{equation}
    \psi(x_b,t_b)\simeq e^{\frac{i}{\hbar}S_{\mathrm{cl}}}\,\psi(x_a,t_a),
    \label{eq:result}
\end{equation}
where $S_{\mathrm{cl}}$ is the action evaluated along the classical trajectory connecting the two events. Eq. \eqref{eq:result} indicates that the wave function at the point $(x_b,t_b)$ can be approximated by the wave function at the point $(x_a,t_a)$ multiplied by a phase factor associated with the classical action, $e^{iS_{\mathrm{cl}}/\hbar}$. Thus, the phase associated with the atomic trajectory 
is 
\begin{equation*}
\Delta_{\text{path}}=\frac{S_{\mathrm{cl}}}{\hbar}.
\end{equation*}

\end{appendices}

\newpage
\bibliography{sn-bibliography}


\end{document}